\documentclass[letterpaper, 10 pt, conference]{ieeeconf}   
\IEEEoverridecommandlockouts                               
\usepackage{optidef}

\usepackage{amssymb,amsmath,color,subfigure}
\usepackage{graphicx}
\usepackage{xspace,stackrel,hyperref}
\usepackage{wrapfig}
\usepackage{mathtools}
\usepackage{tikz,pgfplots}
\usepackage{etoolbox}
\usetikzlibrary{external} 
\usepackage{autonum}
\usepackage{dsfont}
\usepackage{etoolbox}

\usepackage{multirow}
\usepackage{cite}

\usepackage{algorithm, algpseudocode}
\usetikzlibrary{arrows.meta}
\usetikzlibrary{external}

\usepgfplotslibrary{fillbetween}
\usepgfplotslibrary{groupplots}
\usetikzlibrary{decorations.pathreplacing}

\usepackage{lipsum,adjustbox}

\definecolor{mycolor4}{RGB}{230,97,1}
\definecolor{mycolor2}{RGB}{178,171,210}
\definecolor{mycolor3}{RGB}{253,184,99}
\definecolor{mycolor1}{RGB}{94,60,153}

\makeatletter 
\pretocmd\@bibitem{\color{black}\csname keycolor#1\endcsname}{}{\fail}
\newcommand\citecolor[1]{\@namedef{keycolor#1}{\color{blue}}}
\makeatother

\DeclareMathAlphabet{\mathcal}{OMS}{cmsy}{m}{n}

\def\beq{\begin{equation}}
\def\eeq{\end{equation}}

\newcommand{\mc}{\mathcal}

\newcommand{\Z}{\mathbb{Z}}

\newcommand{\R}{\mathds{R}}
\newcommand{\N}{\mathbb{N}}

\newcommand{\defineas}{\coloneqq}

\newcommand{\norm}[1]{\left\lVert#1\right\rVert}

\definecolor{mycolor1}{RGB}{230,97,1}
\definecolor{mycolor2}{RGB}{178,171,210}
\definecolor{mycolor3}{RGB}{253,184,99}
\definecolor{mycolor4}{RGB}{94,60,153}
\definecolor{mycolor5}{rgb}{0,0,0}

\tikzset{
  pics/car/.style args={#1}{
     code={
     \begin{scope}[scale=0.15]
      \shade[top color=#1, bottom color=white, shading angle={135}]
        [draw=black,fill=red!20,rounded corners=0.2ex] (1.5,.5) -- ++(0,1) -- ++(1,0.3) --  ++(3,0) -- ++(1,0) -- ++(0,-1.3) -- (1.5,.5) -- cycle;
    \draw[ rounded corners=0.5ex,fill=black!20!blue!20!white]  (2.5,1.8) -- ++(1,0.7) -- ++(1.6,0) -- ++(0.6,-0.7) -- (2.5,1.8);
    \draw[thick]  (4.2,1.8) -- (4.2,2.5);
    \draw[draw=black,fill=gray!50,thick] (2.75,.5) circle (.5);
    \draw[draw=black,fill=gray!50,thick] (5.5,.5) circle (.5);
    \end{scope}
     }
  }
}

\newtheorem{assumption}{Assumption}

\newtheorem{theorem}{Theorem}

\newtheorem{definition}{Definition}
\newtheorem{lemma}{Lemma}

\newtoggle{full_version}
\toggletrue{full_version}

\title{\LARGE \bf
Receding-Horizon Games with Tullock-Based Profit Functions for Electric Ride-Hailing Markets
}

\author{Marko Maljkovic, Gustav Nilsson, and Nikolas Geroliminis
\thanks{M.~Maljkovic, G.~Nilsson, and N.~Geroliminis are with the School of Architecture, Civil and Environmental Engineering, École Polytechnique Fédérale de Lausanne (EPFL), 1015 Lausanne, Switzerland. {\tt\small \{marko.maljkovic, gustav.nilsson, nikolas.geroliminis\}@epfl.ch}.}%
\thanks{This work was supported by the Swiss National Science Foundation under NCCR Automation, grant agreement 51NF40\_180545.}
\iftoggle{full_version}{}{\thanks{An extended version containing all the proofs is available at \url{http://arxiv.org/abs/2203.09327???}}}%
}

\begin{document}

\maketitle
\thispagestyle{empty}
\pagestyle{empty}

\begin{abstract}
This paper proposes a receding-horizon, game-theoretic charging planning mechanism for electric ride-hailing markets. As the demand for ride-hailing services continues to surge and governments advocate for stricter environmental regulations, integrating electric vehicles into these markets becomes inevitable. The proposed framework addresses the challenges posed by dynamic demand patterns, fluctuating energy costs, and competitive dynamics inherent in such markets. Leveraging the concept of receding-horizon games, we propose a method to optimize proactive dispatching of vehicles for recharging over a predefined time horizon. We integrate a modified Tullock contest that accounts for customer abandonment due to long waiting times to model the expected market share, and by factoring in the demand-based electricity charging, we construct a game capturing interactions between two companies over the time horizon. For this game, we first establish the existence and uniqueness of the Nash equilibrium and then present a semi-decentralized, iterative method to compute it. Finally, the method is evaluated in an open-loop and a closed-loop manner in a simulated numerical case study.
\end{abstract}

\section{Inroduction}

With the ever-increasing spectra of services provided by ride-hailing companies, the ride-hailing markets have arguably solidified their position as transformative forces in urban transportation. Moreover, as the demand for their services continues to grow and governments worldwide push for stricter environmental regulations, it is not unlikely that ride-hailing markets operating electric vehicles (EVs) will play a significant role in shaping the future of urban mobility. Though we are already experiencing a rapid adoption of EVs~\cite{iea}, integrating them in the so-called electric ride-hailing markets characterized by dynamic demand patterns, fluctuating energy costs, and competitive market dynamics, poses various challenges in ensuring company's operational efficiency and maximizing its profitability~\cite{9318522,article4}. As company fleets continue to rise, strategic charging scheduling will become a crucial aspect of their operational management, necessitating considerations of battery discharge rates, demand forecasting, and charging price predictions. As illustrated in Figure~\ref{fig:problem}, a ride-hailing company will aim to be proactive and optimize the number of vehicles dispatched for recharging at each time interval within a predefined horizon. In essence, given that the service demand and charging prices are time-dependent, a company might decide to charge some of the vehicles earlier, before reaching a very low battery level, so as to be able to claim a larger market share during the high demand period. While strategically anticipating forthcoming demand and charging price fluctuations, considering that vehicles dispatched for recharging cannot serve the demand, the optimal charging schedule will primarily be shaped by the expected market share the company can secure at each time interval, all while taking into account the temporal evolution of the aggregate battery level of the fleet. The primary goal of the company is to maximize its profit, defined as the difference between the revenue generated from providing ride-hailing services in a competitive market and the total charging costs incurred over the time horizon. 
\iftoggle{full_version}{\begin {figure}
\centering
\begin{adjustbox}{max height=0.55\textwidth, max width=0.48\textwidth}
\begin{tikzpicture}[scale=0.9]

    \node (pic2) at (-3.0, 0.0) {\includegraphics[width=.09\textwidth]{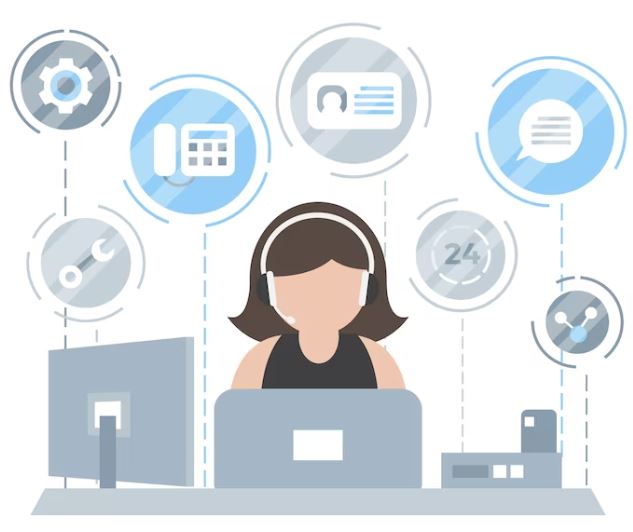}};
    \draw[draw=mycolor1, rounded corners,top color=mycolor1, bottom color=mycolor1, shading angle={90},fill opacity=0.2] (-4.5, 1.5) rectangle (-1.5,-1.5);
    \node[ text=black, shape=rectangle,scale=1.25](ch1) at (-3.0, 1.2){\textbf{Company $a$}};

    \node (pic1) at (3.0, 0.0) {\includegraphics[width=.09\textwidth]{figures/operator.png}};
    \draw[draw=mycolor2, rounded corners,top color=mycolor2, bottom color=mycolor2, shading angle={90},fill opacity=0.2] (1.5, 1.5) rectangle (4.5,-1.5);
    \node[text=black, shape=rectangle,scale=1.25](ch2) at (3.0, 1.2){\textbf{Company $b$}};
    
    \node[scale=1.25] (ba) at (-3.0, -1.2){$\text{X}_a$ vehicles};
    \node[scale=1.25] (bb) at (3.0, -1.2){$\text{X}_b$ vehicles};

    \draw[-{Triangle[length=5pt, width=4.5pt]}, line width=0.5pt] (-3,1.5) -- (-3,2.0)--(3,2)--(3,1.5) ;
    \draw[-{Triangle[length=5pt, width=4.5pt]}, line width=0.5pt] (3,-1.5) -- (3,-2.0)--(-3,-2)--(-3,-1.5) ;


    \draw[-{Triangle[length=5pt, width=4.5pt]}, line width=0.5pt] (-6.5, -5.5)--(6.5, -5.5);
    \draw[ line width=0.5pt] (-1, -5.5)--(-1, -2.5)--(4,-2.5)--(4,-3)--(-1,-3);
    \draw[ line width=0.5pt] (0,  -5.5)--(0, -3.5)--(5,-3.5)--(5,-4)--(0,-4);
    \draw[ line width=0.5pt] (1,  -5.5)--(1, -4.5)--(6,-4.5)--(6,-5)--(1,-5);

    \node[ scale=1](do1) at (-0.5, -2.75) {DO};
    \node[ scale=1](do2) at (0.5, -3.75) {DO};
    \node[ scale=1](do3) at (1.5, -4.75) {DO};
    
    \node[ scale=1](po1) at (2, -2.75) {PLAN};
    \node[ scale=1](po2) at (3, -3.75) {PLAN};
    \node[ scale=1](po3) at (4, -4.75) {PLAN};


    \node[ scale=1.25](k) at (0, -5.85) {$k$};
    \node[ scale=1.25](k1) at (-1, -5.85) {$k-1$};
    \node[scale=1.25](k2) at (1, -5.85) {$k+1$};
    \node[ scale=1.25](t) at (-5.5, -5.85) {Time steps:};
    \node[ scale=2](t) at (-3, -5.85) {$\cdots$};
    \node[ scale=2](t) at (3, -5.85) {$\cdots$};
    \node[ scale=1.25](t) at (5, -5.85) {$k+T$};

    \draw[line width=0.5pt] (0,  -3)--(0, -2.5);
    \draw[line width=0.5pt] (1,  -4)--(1, -3.5);
    \draw[line width=0.5pt] (2,  -5)--(2, -4.5);

    \draw[decorate, decoration={brace, amplitude=5pt, mirror}, line width=0.5pt] (0,-6.5) -- (5,-6.5);
    \draw[line width=0.5pt] (0,  -5.5)--(0, -5.6);
    \draw[line width=0.5pt] (1,  -5.5)--(1, -5.6);
    \draw[line width=0.5pt] (5,  -5.4)--(5, -5.6);
    \draw[line width=0.5pt] (-1,  -5.5)--(-1, -5.6);
    \node[ scale=1.25](t) at (2.5, -7.1) {Planning horizon $T$};

    \draw[rounded corners] (-7, 2.5) rectangle (7,-7.5);
    \node[ scale=1.5, align=center]() at (0, -8.5) {\textbf{At time step $k$, solve charging planning game }\\ \textbf{over a fixed time horizon $T$}};
    
    
    \draw[-{Triangle[length=5pt, width=4.5pt]}, line width=0.5pt] (7,0.5) -- (11,0.5) ;
    \draw[-{Triangle[length=5pt, width=4.5pt]}, line width=0.5pt] (7,-5.5) -- (11,-5.5) ;
    \node[scale = 1.25](veh1) at (8.05,1.0) {\begin{tikzpicture}\draw (0,0) pic{car=mycolor1}; \end{tikzpicture}};
    \node[scale = 1.25](veh2) at (9.55,1.0) {\begin{tikzpicture}\draw (0,0) pic{car=mycolor2}; \end{tikzpicture}};
    \node[scale = 1.25](veh3) at (8.05,-5.0) {\begin{tikzpicture}\draw (0,0) pic{car=mycolor1}; \end{tikzpicture}};
    \node[scale = 1.25](veh4) at (9.55,-5.0) {\begin{tikzpicture}\draw (0,0) pic{car=mycolor2}; \end{tikzpicture}};    

    \draw[rounded corners] (11, 3) rectangle (19,-2);
    \node[ scale=1.5, align=center]() at (15, 3.5) {\textbf{Serve the ride-hailing demand}};
    \draw[rounded corners] (11, -3) rectangle (19,-8);
    \node[ scale=1.5, align=center]() at (15, -8.5) {\textbf{Dispatched for charging}};

    \node [scale = 1](market) at (15,0.5) {\includegraphics[width=.35\textwidth]{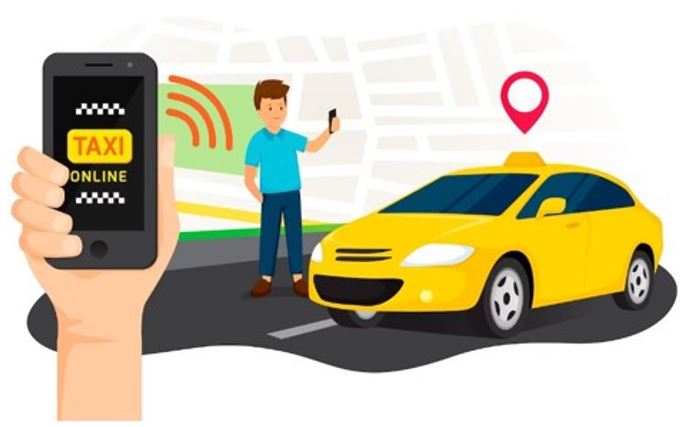}};
    \node [scale = 1](charge) at (15,-5.5) {\includegraphics[width=.35\textwidth]{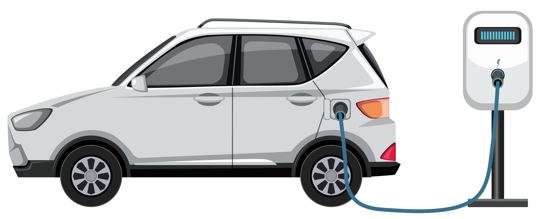}};

    \draw[line width=0.5pt] (0,  -3)--(0, -2.5);


    \draw[-{Triangle[length=5pt, width=4.5pt]},line width=0.5pt] (19, 0.5)--(20.5, 0.5)--(20.5,-2.5);
    \draw[-{Triangle[length=5pt, width=4.5pt]},line width=0.5pt] (19, -5.5)--(20.5, -5.5)--(20.5,-2.5);
    \draw[-{Triangle[length=5pt, width=4.5pt]},line width=0.5pt] (20.5,-2.5)--(21.5,-2.5)--(21.5,-10)--(-9.5,-10)--(-9.5,-5.5)--(-7,-5.5);

    \draw[-{Triangle[length=5pt, width=4.5pt]},line width=0.8pt](19.2,2.2)--(20.8,0.8);
    \node [scale = 0.4](bat1) at (20,1.5) {\includegraphics[width=.1\textwidth]{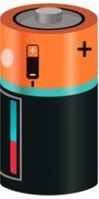}};    

    \draw[-{Triangle[length=5pt, width=4.5pt]},line width=0.8pt](19.2,-7.2)--(20.8,-5.8);
    \node [scale = 0.4](bat1) at (20,-6.5) {\includegraphics[width=.1\textwidth]{figures/batery.jpg}};


    \draw[rounded corners] (-7,4) rectangle (7,9);
    
    \draw[-{Triangle[length=5pt, width=4.5pt]},line width=0.8pt](-7,6.5)--(-9.5,6.5)--(-9.5,0.5)--(-7,0.5);
    \node [scale = 1.5]() at (0,9.5) {\textbf{Demand and charging price forecasting}};

    \draw[-{Triangle[length=5pt, width=4.5pt]},line width=0.8pt](-6,4.75)--(-0.5,4.75);
    \draw[-{Triangle[length=5pt, width=4.5pt]},line width=0.8pt](0.5,4.75)--(6,4.75);

    \draw[-{Triangle[length=5pt, width=4.5pt]},line width=0.8pt](-6,4.75)--(-6,8.25);
    \draw[-{Triangle[length=5pt, width=4.5pt]},line width=0.8pt](0.5,4.75)--(0.5,8.25);

    \node[rotate=90, scale=1.25](n2) at (-6.5, 6.5) {Demand};
    \node[rotate=90, scale=1.25](n2) at (0, 6.5) {Charging price};

    \node[ scale=1.25]() at (-6, 4.4) {$k$};
    \node[ scale=1.25]() at (-1.5, 4.4) {$k+T$};
    \draw[line width=0.5pt](-1.5,4.65)--(-1.5,4.85);

    \node[ scale=1.25]() at (0.5, 4.4) {$k$};
    \node[ scale=1.25]() at (5, 4.4) {$k+T$};
    \draw[line width=0.5pt](5,4.65)--(5,4.85);

    \draw[line width=0.5pt](-6,5)--(-5.1,5)--(-5.1,6)--(-4.2,6)--(-4.2,7.5)--(-3.3,7.5)--(-3.3,6.55)--(-2.4,6.55)--(-2.4,5)--(-1.5,5);

    \draw[line width=0.5pt](0.5,7)--(1.4,7)--(1.4,5.5)--(2.3,5.5)--(2.3,5)--(3.2,5)--(3.2,7)--(4.1,7)--(4.1,7.5)--(5,7.5);
    
{\tiny }
\end{tikzpicture}
\end{adjustbox}
    \caption{The charging planning problem involves two companies solving a scheduling problem at each time step $k$, considering the current battery state of their fleet, forecasted ride-hailing demand, and charging prices over a time horizon $T$.\iftoggle{full_version}{ They can either implement the schedule in an open-loop manner or create a closed-loop system by only executing the first element of the control trajectory and repeating the process.}{}} 
\label{fig:problem}
\end{figure}}{
\begin {figure}
\centering
\begin{adjustbox}{max height=0.28\textwidth, max width=0.48\textwidth}
\begin{tikzpicture}[scale=0.9]

    \node (pic2) at (-3.0, 0.0) {\includegraphics[width=.09\textwidth]{figures/operator.png}};
    \draw[draw=mycolor1, rounded corners,top color=mycolor1, bottom color=mycolor1, shading angle={90},fill opacity=0.2] (-4.5, 1.5) rectangle (-1.5,-1.5);
    \node[ text=black, shape=rectangle,scale=1.25](ch1) at (-3.0, 1.2){\textbf{Company $a$}};

    \node (pic1) at (3.0, 0.0) {\includegraphics[width=.09\textwidth]{figures/operator.png}};
    \draw[draw=mycolor2, rounded corners,top color=mycolor2, bottom color=mycolor2, shading angle={90},fill opacity=0.2] (1.5, 1.5) rectangle (4.5,-1.5);
    \node[text=black, shape=rectangle,scale=1.25](ch2) at (3.0, 1.2){\textbf{Company $b$}};
    
    \node[scale=1.25] (ba) at (-3.0, -1.2){$\text{X}_a$ vehicles};
    \node[scale=1.25] (bb) at (3.0, -1.2){$\text{X}_b$ vehicles};

    \draw[-{Triangle[length=5pt, width=4.5pt]}, line width=0.5pt] (-3,1.5) -- (-3,2.0)--(3,2)--(3,1.5) ;
    \draw[-{Triangle[length=5pt, width=4.5pt]}, line width=0.5pt] (3,-1.5) -- (3,-2.0)--(-3,-2)--(-3,-1.5) ;


    \draw[-{Triangle[length=5pt, width=4.5pt]}, line width=0.5pt] (-6.5, -5.5)--(6.5, -5.5);
    \draw[ line width=0.5pt] (-1, -5.5)--(-1, -2.5)--(4,-2.5)--(4,-3)--(-1,-3);
    \draw[ line width=0.5pt] (0,  -5.5)--(0, -3.5)--(5,-3.5)--(5,-4)--(0,-4);
    \draw[ line width=0.5pt] (1,  -5.5)--(1, -4.5)--(6,-4.5)--(6,-5)--(1,-5);

    \node[ scale=1](do1) at (-0.5, -2.75) {DO};
    \node[ scale=1](do2) at (0.5, -3.75) {DO};
    \node[ scale=1](do3) at (1.5, -4.75) {DO};
    
    \node[ scale=1](po1) at (2, -2.75) {PLAN};
    \node[ scale=1](po2) at (3, -3.75) {PLAN};
    \node[ scale=1](po3) at (4, -4.75) {PLAN};


    \node[ scale=1.25](k) at (0, -5.85) {$k$};
    \node[ scale=1.25](k1) at (-1, -5.85) {$k-1$};
    \node[scale=1.25](k2) at (1, -5.85) {$k+1$};
    \node[ scale=1.25](t) at (-5.5, -5.85) {Time steps:};
    \node[ scale=2](t) at (-3, -5.85) {$\cdots$};
    \node[ scale=2](t) at (3, -5.85) {$\cdots$};
    \node[ scale=1.25](t) at (5, -5.85) {$k+T$};

    \draw[line width=0.5pt] (0,  -3)--(0, -2.5);
    \draw[line width=0.5pt] (1,  -4)--(1, -3.5);
    \draw[line width=0.5pt] (2,  -5)--(2, -4.5);

    \draw[decorate, decoration={brace, amplitude=5pt, mirror}, line width=0.5pt] (0,-6.5) -- (5,-6.5);
    \draw[line width=0.5pt] (0,  -5.5)--(0, -5.6);
    \draw[line width=0.5pt] (1,  -5.5)--(1, -5.6);
    \draw[line width=0.5pt] (5,  -5.4)--(5, -5.6);
    \draw[line width=0.5pt] (-1,  -5.5)--(-1, -5.6);
    \node[ scale=1.25](t) at (2.5, -7.1) {Planning horizon $T$};

    \draw[rounded corners] (-7, 2.5) rectangle (7,-7.5);
    \node[ scale=1.5, align=center]() at (0, -8.5) {\textbf{At time step $k$, solve charging planning game }\\ \textbf{over a fixed time horizon $T$}};
    
    
    \draw[-{Triangle[length=5pt, width=4.5pt]}, line width=0.5pt] (7,0.5) -- (11,0.5) ;
    \draw[-{Triangle[length=5pt, width=4.5pt]}, line width=0.5pt] (7,-5.5) -- (11,-5.5) ;
    \node[scale = 1.25](veh1) at (8.05,1.0) {\begin{tikzpicture}\draw (0,0) pic{car=mycolor1}; \end{tikzpicture}};
    \node[scale = 1.25](veh2) at (9.55,1.0) {\begin{tikzpicture}\draw (0,0) pic{car=mycolor2}; \end{tikzpicture}};
    \node[scale = 1.25](veh3) at (8.05,-5.0) {\begin{tikzpicture}\draw (0,0) pic{car=mycolor1}; \end{tikzpicture}};
    \node[scale = 1.25](veh4) at (9.55,-5.0) {\begin{tikzpicture}\draw (0,0) pic{car=mycolor2}; \end{tikzpicture}};    

    \draw[rounded corners] (11, 3) rectangle (19,-2);
    \node[ scale=1.5, align=center]() at (15, 3.5) {\textbf{Serve the ride-hailing demand}};
    \draw[rounded corners] (11, -3) rectangle (19,-8);
    \node[ scale=1.5, align=center]() at (15, -8.5) {\textbf{Dispatched for charging}};

    \node [scale = 1](market) at (15,0.5) {\includegraphics[width=.35\textwidth]{figures/Market.jpg}};
    \node [scale = 1](charge) at (15,-5.5) {\includegraphics[width=.35\textwidth]{figures/ch1.jpg}};

    \draw[line width=0.5pt] (0,  -3)--(0, -2.5);


    \draw[-{Triangle[length=5pt, width=4.5pt]},line width=0.5pt] (19, 0.5)--(20.5, 0.5)--(20.5,-2.5);
    \draw[-{Triangle[length=5pt, width=4.5pt]},line width=0.5pt] (19, -5.5)--(20.5, -5.5)--(20.5,-2.5);
    \draw[-{Triangle[length=5pt, width=4.5pt]},line width=0.5pt] (20.5,-2.5)--(21.5,-2.5)--(21.5,-10)--(-9.5,-10)--(-9.5,-5.5)--(-7,-5.5);

    \draw[-{Triangle[length=5pt, width=4.5pt]},line width=0.8pt](19.2,2.2)--(20.8,0.8);
    \node [scale = 0.4](bat1) at (20,1.5) {\includegraphics[width=.1\textwidth]{figures/batery.jpg}};    

    \draw[-{Triangle[length=5pt, width=4.5pt]},line width=0.8pt](19.2,-7.2)--(20.8,-5.8);
    \node [scale = 0.4](bat1) at (20,-6.5) {\includegraphics[width=.1\textwidth]{figures/batery.jpg}};


    \draw[rounded corners] (-7,4) rectangle (7,9);
    
    \draw[-{Triangle[length=5pt, width=4.5pt]},line width=0.8pt](-7,6.5)--(-9.5,6.5)--(-9.5,0.5)--(-7,0.5);
    \node [scale = 1.5]() at (0,9.5) {\textbf{Demand and charging price forecasting}};

    \draw[-{Triangle[length=5pt, width=4.5pt]},line width=0.8pt](-6,4.75)--(-0.5,4.75);
    \draw[-{Triangle[length=5pt, width=4.5pt]},line width=0.8pt](0.5,4.75)--(6,4.75);

    \draw[-{Triangle[length=5pt, width=4.5pt]},line width=0.8pt](-6,4.75)--(-6,8.25);
    \draw[-{Triangle[length=5pt, width=4.5pt]},line width=0.8pt](0.5,4.75)--(0.5,8.25);

    \node[rotate=90, scale=1.25](n2) at (-6.5, 6.5) {Demand};
    \node[rotate=90, scale=1.25](n2) at (0, 6.5) {Charging price};

    \node[ scale=1.25]() at (-6, 4.4) {$k$};
    \node[ scale=1.25]() at (-1.5, 4.4) {$k+T$};
    \draw[line width=0.5pt](-1.5,4.65)--(-1.5,4.85);

    \node[ scale=1.25]() at (0.5, 4.4) {$k$};
    \node[ scale=1.25]() at (5, 4.4) {$k+T$};
    \draw[line width=0.5pt](5,4.65)--(5,4.85);

    \draw[line width=0.5pt](-6,5)--(-5.1,5)--(-5.1,6)--(-4.2,6)--(-4.2,7.5)--(-3.3,7.5)--(-3.3,6.55)--(-2.4,6.55)--(-2.4,5)--(-1.5,5);

    \draw[line width=0.5pt](0.5,7)--(1.4,7)--(1.4,5.5)--(2.3,5.5)--(2.3,5)--(3.2,5)--(3.2,7)--(4.1,7)--(4.1,7.5)--(5,7.5);
    
{\tiny }
\end{tikzpicture}
\end{adjustbox}
    \caption{The charging planning problem involves two companies solving a scheduling problem at each time step $k$, considering the current battery state of their fleet, forecasted ride-hailing demand, and charging prices over a time horizon $T$.\iftoggle{full_version}{ They can either implement the schedule in an open-loop manner or create a closed-loop system by only executing the first element of the control trajectory and repeating the process.}{}} 
\label{fig:problem}
\vspace{-2em}
\end{figure}
}

Game theory has emerged as a promising tool for tackling the EV charging problem~\cite{EVsCharging,9102356,article5,10383725} and integrating it with the intricate dynamics of ride-hailing markets~\cite{10384201,Hierarchical, ECC2023, maljkovic2023learning, ECC2024}. However, existing studies predominantly focus on static setups, meaning that they model interactions between companies and compute the equilibrium for only one snapshot of the day. In contrast, dynamic mechanisms, taking into account competing interactions over a certain time period, are common in the literature on demand-side management (DSM). In terms of dynamic modeling, the framework presented in this paper shares similarities with the ones presented in~\cite{9992497, 6994293, 9354436, ESTRELLA2019126}, wherein authors introduce the concepts of receding-horizon games (RHG) inspired by the well-established paradigm of model predictive control (MPC). Nevertheless, the game played over the horizon differs significantly from the one analyzed in this paper, as they primarily focus on quadratic or aggregative games, for which there exists an extensive body of literature on efficiently computing the Nash equilibrium.

In this paper, we propose a receding-horizon, game-theoretic charging planning mechanism for a market comprising two ride-hailing companies.  Specifically, for each company, we adopt a simplified aggregate model of battery discharging, resulting in a linear state-space model. Under this model, and similar to~\cite{ECC2024}, we treat the expected market share from serving the ride-hailing demand at each time interval within the planning horizon as a modified Tullock contest~\cite{OSORIO2013164, KIM20181} that integrates information about customer abandonment. This is different from existing literature that incorporates standard Tullock contests in a sense that companies will not claim the whole profit that can be earned during a particular time interval, but rather a certain portion of it will be forfeited due to customers canceling their service requests owing to prolonged waiting times. To the best of our knowledge, in the context of receding-horizon games, this framework is the first one to integrate a Tullock-based expected revenue model with demand-driven charging costs to characterize players' profits at each time interval within the horizon. From the theoretical perspective, one of the main contributions of this paper encompasses establishing both the existence and uniqueness of the Nash equilibrium for the game played across the predefined horizon. Additionally, we present an iterative algorithm with provable convergence guarantees for computing the game's Nash equilibrium in a semi-decentralized manner. In terms of practical applicability, we design a simulated case study to evaluate the performance of the proposed method. We assess it both when applied in an open-loop fashion, where the planning horizon matches the length of the analyzed time frame, and in a closed-loop, receding-horizon fashion, where the planning horizon is shorter than the length of the analyzed time frame.

The paper is outlined as follows: the rest of this section introduces some basic notation. In Section~\ref{sec:model}, we present the market model and the general problem statement. Then, in Section~\ref{sec:NE}, we outline our main methodological and theoretical results and conclude with Sections~\ref{sec:example} and~\ref{sec:conclusion}, where we illustrate the performance of our method in a numerical case study and propose ideas for future research.

\textit{Notation:}  Let $\R_{(+)}$ and $\Z_{(+)}$ denote the sets of (non-negative) real and integer numbers. Let $\mathbf{1}_{m}$ and $\mathbf{I}_m$ be the vector of all ones and a unit matrix of size $m$. For any $T\in\Z_+$, we let $\Z_T=\{0,1,2,...,T-1\}$. If $\mc A$ is a finite set of vectors $x_i$, we define $x \defineas \text{col}((x_{i})_{i\in \mc A})$ to be their concatenation. For a vector $x\in\R^n$, we let $\text{diag}(x)\in\R^{n\times n}$ denote a diagonal matrix whose elements on the diagonal correspond to vector $x$. If $\mc A$ is a set of $k$ matrices $A_i\in\R^{m\times n}$, then $\text{blkdiag}(\{A_i\}_{i\in\mc A})\in\R^{km\times kn}$ denotes the corresponding block-diagonal matrix and $\text{vstack}((A_i)_{i\in\mc A})\in\R^{km\times n}$ denotes their vertical concatenation.

\section{Model}\label{sec:model}
We consider a ride-hailing market comprising two companies $i\in\{a,b\}$, operating fleets of $\text{X}_a$ and $\text{X}_b$ electric vehicles respectively. Over a horizon of $T\in\N$ future time intervals, these companies acquire estimates of relevant exogenous parameters and aim to optimize the management of their electric fleets. At every time interval $k\in\Z_T$, the aggregate battery state of company $i$'s fleet is described by a vector $x_i[k]\defineas\text{col}((x^j_i[k])_{j\in\mc B})\in\R_+^m$, where $m\in\N$ denotes the number of distinct battery level categories and $\mc B=\Z_m$. Namely, we assume that for all $k\in\Z_T$, each of the $\text{X}_i$ vehicles belongs to one of the $m$ categories given its current battery level, that $x_i^j[k]\geq 0$ denotes the total number of vehicles in category $j\in\mc B$, and that $\mathbf{1}_m^T x_i[k]=\text{X}_i$. 

Vehicles can be categorized into three groups based on their status: those currently in operation, those dispatched for charging, and those temporarily parked. During each time interval $k\in\Z_T$, every company determines the quantity of vehicles from each category $j\in\mc B$, i.e., $u_i^j[k]\in\R_+$, to dispatch for recharging, i.e., every company $i\in\{a,b\}$ chooses a vector-valued control input $u_i[k]\defineas\text{col}((u_i^j[k])_{j\in\mc B})\in\R_+^m$. To describe the state of charge on the aggregate level of the fleet, we adopt the following simplifying assumptions:
\begin{enumerate}
        \item Vehicles dispatched for charging are unavailable to serve the ride-hailing demand during the respective interval $k\in\Z_T$. The ones dispatched from categories $\mc B\setminus \{m-1\}$ transit from the current category to the adjacent higher one at the end of the interval, whereas the ones from category $m-1$ remain in the same one.
        \item Battery state of the vehicles on the lowest level, i.e., level $0$, is deemed critical so they cannot serve the demand. They either have to be parked or sent to recharge during the particular interval $k\in\Z_T$.
        \item Undispatched vehicles from categories $\mc B\setminus \{0\}$ serve the ride-hailing demand in the region during the respective interval. For every company $i$ and every battery level category $j\in\mc B$, $\alpha^j_i\in[0,1]$ determines the portion of undispatched vehicles from category $j$ that will remain in the same category at the end of the interval while the remaining ones transit to the next lower category.  
\end{enumerate}
\iftoggle{full_version}{With this in mind, for every company $i\in\{a,b\}$, we model the dynamics of the fleet's state of charge as $x_i[k+1]=\text{col}((f_i^j(x_i[k],u_i[k]))_{j\in\mc B})$, where $f_i^j:\R^m_+\times\R^m_+\rightarrow\R_+$ is defined by following equations:

\noindent- If $j=m-1$, then $f_i^j(x_i[k],u_i[k])=\alpha_i^j(x_i^j[k]-u^j_i[k])+u_i^j[k]+u_i^{j-1}[k]$.

\medskip
\noindent- If $j\in\mc B\setminus\{0,m-1\}$, then $f_i^j(x_i[k],u_i[k])=\alpha_i^j(x_i^j[k]-u^j_i[k])+(1-\alpha_i^{j+1})(x_i^{j+1}[k]-u^{j+1}_i[k])+u_i^{j-1}[k]$.

\medskip
\noindent- If $j=0$, then $f_i^j(x_i[k],u_i[k])=x_i^j[k]-u_i^{j}[k]+(1-\alpha_i^{j+1})(x_i^{j+1}[k]-u^{j+1}_i[k])$.

\medskip}{With this in mind, for every company $i\in\{a,b\}$, we model the dynamics of the fleet's state of charge as $x_i[k+1]=\text{col}((f_i^j(x_i[k],u_i[k]))_{j\in\mc B})$ for particularly chosen  $f_i^j:\R^m_+\times\R^m_+\rightarrow\R_+$.}
\noindent If the initial state of the fleet, i.e., $x_i[0]=x_i^0$ is known, then these conditions form a linear state-space model
\begin{equation}
    \label{eq:1}
        x_i[k+1]=A_ix_i[k]+B_iu_i[k]\,,
\end{equation}
with $A_i\defineas A_i(\{\alpha_i^j\}_{j\in\mc B})$ and $B_i\defineas B_i(\{\alpha_i^j\}_{j\in\mc B})$ being adequately chosen square matrices. By introducing the vector $\Lambda^T=[1,\dots,1,0]$ of size $m$, we can establish company $i$'s count of operating vehicles at $k\in\Z_T$, i.e., $\phi_i[k]\in\R_+$, as
\begin{equation}
    \label{eq:2}
    \phi_i[k]=\Lambda^T(x_i[k]-u_i[k])\,,
\end{equation}
and consequently form $\mathbf{\Phi}_i\defineas\text{col}((\phi_i[k])_{k\in\Z_T})$ and $\mathbf{U}_i\defineas\text{col}((u_i[k])_{k\in\Z_T})$ describing company $i$'s operation over the whole time horizon. In the same manner, we define for the rival company $\phi_{-i}[k]\in\R_+$, $\mathbf{\Phi}_{-i}\defineas\text{col}((\phi_{-i}[k])_{k\in\Z_T})$, $u_{-i}[k]\in\R_+^m$, and $\mathbf{U}_{-i}\defineas\text{col}((u_{-i}[k])_{k\in\Z_T})$ that depict the number of operating and dispatched vehicles. Finally, it is important to note that for the dynamics to be feasible, 
\begin{equation}
    \label{eq:3}
    0\leq u_i^j[k]\leq x_i^j[k]\,
\end{equation}
needs to hold for every $i\in\{a,b\}$, $j\in\mc B$, and $k\in\Z_T$.\iftoggle{full_version}{

Having formally introduced the ride-hailing fleet model, we can now continue to present the problem formulation.}{} 

\subsection{Problem statement}\label{subsec:ps}

The goal of every ride-hailing company is to optimize their profit within the time horizon $\Z_T$, by strategically determining the number of vehicles to dispatch for charging at each time interval. If we let $\mathbf{Z}_i\in\R_+^{T+mT}$ be given by $\mathbf{Z}_i\defineas[\mathbf{\Phi}_i^T,\:\:\mathbf{U}_i^T]^T$, 
then, we can model the profit of each company $i\in\{a,b\}$ over the whole time horizon as:
\begin{multline}
\label{eq:4}
    \mathbf{J}_i(\mathbf{Z}_i, \mathbf{Z}_{-i})=  \\ 
\sum_{k\in\Z_T}\mathbf{J}_{i,k}^{\text{market}}(\phi_i[k],\phi_{-i}[k])-\mathbf{J}_{i,k}^{\text{charge}}(u_i[k], u_{-i}[k])\,.
\end{multline}
Here, $\mathbf{J}_{i,k}^{\text{market}}(\phi_i[k],\phi_{-i}[k])$ denotes the expected ride-hailing market share the company can secure at $k\in\Z_T$ and 
$\mathbf{J}_{i,k}^{\text{charge}}(u_i[k], u_{-i}[k])$ represents the average cost of charging for the currently dispatched vehicles. 

The expected revenue generated from serving the ride-hailing market varies over time, as demand typically rises during peak-hour periods. Moreover, time intervals with higher anticipated demand call for a greater number of operating vehicles to reduce the profit loss caused by customer abandonments resulting from extended waiting times for vehicle assignments. Similar to~\cite{ECC2024}, we assume a portion of the revenue from ride-hailing requests will
constantly be forfeited and hence adopt the following market model:
\begin{equation}
    \label{eq:5}
    \mathbf{J}_{i,k}^{\text{market}}(\phi_i[k],\phi_{-i}[k])=n[k]r[k]\frac{\phi_i[k]}{\phi_i[k] + \phi_{-i}[k] + \varepsilon[k]}\,,
\end{equation}
where $n[k]>0$ denotes the average number of requests, $r[k]>0$ denotes the average revenue per request, and $\varepsilon[k]>0$ models the profit loss due to abandonments at $k\in\Z_T$. Specifically, the time distribution of abandonments, given by $\varepsilon[k]>0$, dictates the portion of the total potential profit forfeited over the horizon:
\begin{equation}
    \label{eq:6}
    l_T =\sum_{k\in\Z_T}n[k]r[k]\frac{\varepsilon[k]}{\phi_i[k] + \phi_{-i}[k] + \varepsilon[k]}\,. 
\end{equation}
At every $k\in\Z_T$, it is evident that the profit loss decreases as the total number of vehicles participating in the market at that time interval, i.e., $\phi_i[k] + \phi_{-i}[k]$, increases. Thus, although similar to classical Tullock contests~\cite{OSORIO2013164, KIM20181, LI2022102771} that assume the whole potential profit would be shared between the companies,~\eqref{eq:5} acknowledges that reduced participation in the market leads to diminished profits for both companies.

We adopt a setup where the price of charging during time slot $k\in\mathbb{Z}_T$ for each category $j\in\mathbb{Z}_m$, denoted as $p^j[k]\in\mathbb{R}_+$, is determined by the total charging demand within that particular category, i.e., $p^j[k]=c^j[k]\overline{d}[k](u^j_i[k]+u^j_{-i}[k])$.
Here, $\overline{d}[k]>0$ and $c^j[k]>0$ represent the exogenous parameters depicting the average charging demand per vehicle and the price per unit of energy associated with battery level $j\in\Z_m$. In this context, we perceive $\overline{d}[k]$ as influenced by battery discharge rates of the vehicles and the average vehicle speed in the region that fluctuates throughout the day based on network congestion. Assuming that idle vehicles continue searching for passengers and that their discharge rates are comparable, we simplify the setup by regarding this parameter as company-independent. Finally, the total charging cost at $k\in\Z_T$ is $\mathbf{J}_{i,k}^{\text{charge}}(u_i[k], u_{-i}[k])$ given by 
\begin{equation}
    \label{eq:8}
    \mathbf{J}_{i,k}^{\text{charge}}(u_i[k], u_{-i}[k])=\sum_{j\in\Z_m}p^j[k]\overline{d}[k]u_i^j[k]\,.
\end{equation}
If we let $\beta[k]\defineas n[k]r[k]$, $q^j[k]\defineas c^j[k]\overline{d}^2[k]$, $W[k]\defineas\text{diag}(\{q^j[k]\}_{j\in\Z_m})$, and $\overline{\mathbf{W}}\defineas\text{blkdiag}(\{W[k]\}_{k\in\Z_T})$, then the profit of each company can be compactly written as 
\begin{equation}
    \label{eq:9}
    \mathbf{J}_i(\mathbf{Z}_i, \mathbf{Z}_{-i}) = \textbf{g}_1(\mathbf{\Phi}_i,\mathbf{\Phi}_{-i})+\textbf{g}_2(\mathbf{U}_i,\mathbf{U}_{-i})\,,
\end{equation}
where $\textbf{g}_1(\mathbf{\Phi}_i,\mathbf{\Phi}_{-i})$ and $\textbf{g}_2(\mathbf{U}_i,\mathbf{U}_{-i})$ are given by:
$$\textbf{g}_1(\mathbf{\Phi}_i,\mathbf{\Phi}_{-i})\defineas \sum_{k\in\Z_T}\frac{\beta[k]\phi_i[k]}{\phi_i[k] + \phi_{-i}[k] + \varepsilon[k]}\,,$$
$$\textbf{g}_2(\mathbf{U}_i,\mathbf{U}_{-i})\defineas -\mathbf{U}_i^T\overline{\mathbf{W}}(\mathbf{U}_i+\mathbf{U}_{-i})\,.$$
\medskip
\noindent If for every $i\in\{a,b\}$ we define a set $\mc Z_i\subseteq\R_+^{T+mT}$ as $\mc Z_i\defineas\{\mathbf{Z}_i \mid x_i[0]=x_i^0 \land \text{\eqref{eq:1},~\eqref{eq:2}, and~\eqref{eq:3} hold}\}$, then the interactions between the two companies can be described as a two-player game $\mc G$ defined by a set of coupled best-response optimization problems 
\begin{equation}\label{eq:12}
    \mc G\defineas\bigl\{\max_{\mathbf{Z}_i\in\mc Z_i}\textbf{g}_1(\mathbf{\Phi}_i,\mathbf{\Phi}_{-i})+\textbf{g}_2(\mathbf{U}_i,\mathbf{U}_{-i}),\forall i\in\{a,b\}\bigr\}\,.
\end{equation}
A no-regret solution for the companies is given by a pair of strategies that constitute a Nash equilibrium (NE). Namely, if  $\mathbf{Z}\in\mc Z$ with $\mc Z\defineas \mc Z_a\times\mc Z_b$, denotes the joint strategy of the two companies, i.e., $\mathbf{Z}\defineas\text{col}((\mathbf{Z}_i)_{i\in\{a,b\}})$, then the NE concept can be formally introduced as in Definition~\ref{def:1}.  
\begin{definition}[Nash equilibrium]\label{def:1}
    A joint strategy $\overline{\mathbf{Z}}\in\mc Z$ is a Nash equilibrium of $\mc G$ given by~\eqref{eq:12}, if for all $i\in\{a,b\}$ and all $\mathbf{Z}_i\in\mc Z_i$ it holds that $\mathbf{J}_i(\overline{\mathbf{Z}}_i,\overline{\mathbf{Z}}_{-i})\geq\mathbf{J}_i(\mathbf{Z}_i,\overline{\mathbf{Z}}_{-i})$.
\end{definition}
\medskip
Computing the NE strategy for a two-player, receding-horizon game defined by ride-hailing market~\eqref{eq:12} essentially boils down to solving a charging planning problem for the two companies over the horizon $T$. In that sense, the setup of~\eqref{eq:12} resembles the standard concept of Model Predictive Control (MPC) and can be used to also establish feedback control policies, should the companies wish to optimize expected profit within a time frame $T_{\text{total}}$ that is longer than the planning horizon $T$. Therefore, in the following section we first prove the existence and uniqueness of the NE for the game~$\mc G$ and then proceed to present a method to compute it. 

\section{Characterizing the Nash equilibria}\label{sec:NE}

To begin the analysis, we first direct our attention to the constraint set $\mc Z_i$. In order to reformulate~\eqref{eq:3} as a constraint on $\mathbf{Z}_i$, we recall that the state trajectory of the linear system~\eqref{eq:1} is given by $x_i[k]=A_i^kx_i^0+\sum_{l=0}^{k-1}A_i^{k-l-1}B_iu_i[l]$ for $x_i[0]=x_i^0$. Combining this with the right-hand side of~\eqref{eq:3}, for every $k\geq 0$ we obtain
\begin{equation}
    \label{eq:14}
    \left[A_i^{k-1}B_i | A_i^{k-2}B_i | \dots | A_iB_i | B_i | -\mathbf{I}_m\right]U_i^{0:k}\geq -A_i^kx_i^0\,,
\end{equation}
where the inequality is element-wise and  $U_i^{0:k}\in\R_+^{(k+1)m}$ is the concatenation of control inputs corresponding to time intervals $0\leq l\leq k$, i.e., $U_i^{0:k}\defineas\text{col}((u_i[l])_{l\in\Z_{k+1}})$. For every $k\in\R_+$, let us now define $\text{r}_i[k]\defineas -A_i^kx_i^0$ and
\begin{equation}
    \label{eq:15}
    \text{L}_i[k]=\left[A_i^{k-1}B_i | \dots | A_iB_i | B_i | -\mathbf{I}_{m}|\mathbf{0}_{m\times m(T-k-1)}\right]\,.
\end{equation}
that will help us connect player $i$'s optimization variable with constraints~\eqref{eq:14}. It is clear that company $i$'s control input over the horizon $T$ satisfies $\mathbf{U}_i=U_i^{0:T-1}$. Hence, for $0\leq k\leq T-1$, each constraint~\eqref{eq:14} can be represented as $\text{L}_i[k]\mathbf{U}_i\geq\text{r}_i[k]$. Similarly, for every $k\geq0$, condition~\eqref{eq:2} gives us $\phi_i[k]+\Lambda^T\left[A_i^{k-1}B_i | \dots | A_iB_i | B_i | -\mathbf{I}_m\right]U_i^{0:k}  =-\Lambda^TA_i^kx_i^0$, which for $k$ satisfying $0\leq k\leq T-1$ gives
\iftoggle{full_version}{\begin{equation}
    \label{eq:17}
    -\phi_i[k]+\Lambda^T\text{L}_i[k]\mathbf{U}_i=\Lambda^T\text{r}_i[k]\,.
\end{equation}}{$\phi_i[k]+\Lambda^T\text{L}_i[k]\mathbf{U}_i=\Lambda^T\text{r}_i[k]$.}
If we stack matrices $\text{L}_{i}[k]$ and vectors $\text{r}_i[k]$ for $k\in\Z_T$, i.e.,
\iftoggle{full_version}{\begin{equation}
    \label{eq:18}
    \mathbf{L}_i=\text{vstack}((\text{L}_i[k])_{k\in\Z_T})\:\:\text{and}\:\:\mathbf{r}_i=\text{col}((\text{r}_i[k])_{k\in\Z_T})\,,
\end{equation}}{$\mathbf{L}_i=\text{vstack}((\text{L}_i[k])_{k\in\Z_T})\:\:\text{and}\:\:\mathbf{r}_i=\text{col}((\text{r}_i[k])_{k\in\Z_T})$,}
then the set $\mc Z_i\subseteq\R_+^{T+mT}$ can be described by a polytope 
\begin{equation}
    \label{eq:19}
    \mc Z_i=\{\mathbf{Z}_i\mid \mathbf{L}_{\text{inq},i}\mathbf{Z}_i\leq\mathbf{r}_{\text{inq},i}\:\land\:\mathbf{L}_{\text{eq},i}\mathbf{Z}_i=\mathbf{r}_{\text{eq},i}\}\,,
\end{equation}
with $\mathbf{L}_{\text{inq},i}=\left[\begin{array}{cc}
   \mathbf{0}_{mT\times T}  & -\mathbf{L}_i \\
   \mathbf{0}_{mT\times T}  & -\mathbf{I}_{mT}
\end{array}\right]$, $\mathbf{r}_{\text{inq},i}^T=-[\mathbf{r}_i^T;\:\mathbf{0}_{mT}^T]$,
$
    \mathbf{L}_{\text{eq},i}=[-\mathbf{I}_T;\:(\mathbf{I}_T\otimes\Lambda^T)\mathbf{L}_i]$ and $\mathbf{r}_{\text{eq},i}=(\mathbf{I}_T\otimes\Lambda^T)\mathbf{r}_i$.
\iftoggle{full_version}{By construction, state-space model~\eqref{eq:1} ensures that $\mathbf{1}_m^Tx_i[k+1]=\mathbf{1}_m^Tx_i[k]=\text{X}_i$ holds for every $k\in\Z_T$, which, alongside~\eqref{eq:3}, ensures that polytope $\mc Z_i$ is compact. On the other hand, Slater's constraint qualification has to be examined with care as it depends on the initial condition $x_i[0]=x_i^0$. Namely, if for some $l\in\Z_m$ the corresponding element of the initial state is zero, i.e., $x_i^l[0]=0$, then due to $0\leq u_i^l[0]\leq x_i^l[0]$, we immediately get $u_i^l[0]=0$. Hence, for Slater's condition to hold, the corresponding inequality constraint in~\eqref{eq:19} should be framed as an equality one. Without loss of generality, we can adopt the following assumption.}{By construction,~\eqref{eq:1} and~\eqref{eq:3} ensure that $\mc Z_i$ is compact. Without loss of generality, we adopt the following assumption.}
\begin{assumption}\label{ass:2}
    Let game $\mc G$ be defined as~\eqref{eq:12} with polytopic sets $\mc Z_i$ as in~\eqref{eq:19}. For every $i\in\{a,b\}$, $x_i^0\in\R_+^m$ is chosen such that $\mc Z_i$ satisfies Slater's constraint qualification. 
\end{assumption}
\medskip

Based on the structure of the profit functions~\eqref{eq:9} and constraint sets~\eqref{eq:19}, we can now show that there exists a unique NE of the game $\mc G$ defined in~\eqref{eq:12}.
\begin{theorem}\label{th:1}
    Let game $\mc G$ be defined as~\eqref{eq:12}, with feasible sets $\mc Z_i$ defined by~\eqref{eq:19}. If Assumption~\ref{ass:2} holds, then $\mc G$ admits a unique Nash equilibrium. 
\end{theorem}
\iftoggle{full_version}{\begin{proof}
    We start by observing that $\mathbf{J}_i(\mathbf{Z}_i,\mathbf{Z}_{-i})$ is continuous in $\mathbf{Z}\in\mc Z$. For any given $\mathbf{Z}_{-i}\in\mc Z_{-i}$, the profit functions are concave in $\mathbf{Z}_i\in\mc Z_i$ since $\varepsilon[k]>0,\psi_i[k]\geq0,\psi_{-i}[k]\geq0$, 
   \begin{equation}
        \frac{\partial^2\mathbf{J}_i}{\partial\mathbf{\Phi}_i^2}=\text{diag}\left(\left\{\frac{-2\beta[k](\phi_{-i}[k]+\varepsilon[k])}{(\phi_i[k]+\phi_{-i}[k]+\varepsilon[k])^3}\right\}_{k\in\Z_T}\right)\prec 0\,,
    \end{equation}
    \begin{equation}
        \frac{\partial^2\mathbf{J}_i}{\partial\mathbf{U}_i^2}=-2\overline{\mathbf{W}}\prec 0\:,\:\frac{\partial^2\mathbf{J}_i}{\partial\mathbf{\Phi}_i\partial\mathbf{U}_i}=\mathbf{0}\:\text{and}\:\frac{\partial^2\mathbf{J}_i}{\partial\mathbf{U}_i\partial\mathbf{\Phi}_i}=\mathbf{0}\,,
    \end{equation}
which yields a negative definite diagonal Hessian $\nabla^2_{\mathbf{Z}_i}\mathbf{J}_i$. Assumption~\ref{ass:2}, guarantees that sets $\mc Z_i$ are compact, convex and satisfy Slater's constraint qualification. Therefore, we can directly invoke~\cite[Th.1]{Rosen} to prove the existence of the NE. To prove uniqueness, let $e=\text{col}((e_i)_{i\in\{a,b\}})\in\R^2$ be a 
 vector such that $e_i>0$, and
\begin{equation}
    \label{eq:20}
    g(\mathbf{Z},e)\defineas\text{col}\left((e_{i}\nabla_{\mathbf{Z}_i}\mathbf{J}_i(\mathbf{Z}_i,\mathbf{Z}_{-i}))_{i\in \{a,b\}}\right)\,.
\end{equation}  
Let $G(\mathbf{Z},e)$ denote the Jacobian of $g(\mathbf{Z},e)$ with respect to $\mathbf{Z}$. Then, a sufficient condition for the uniqueness of the Nash equilibrium is that the matrix $\Gamma\defineas G(\mathbf{Z},e)+G^{T}(\mathbf{Z},e)$
be negative definite for all $\mathbf{Z}\in\mc Z$ and some $e\in \R^2_{>0}$~\cite[Th.2]{Rosen}. Considering that
    \begin{equation}
        \frac{\partial^2\mathbf{J}_i}{\partial\mathbf{\Phi}_{-i}\partial\mathbf{\Phi}_i}=\text{diag}\left(\left\{\frac{\beta[k](\phi_{i}[k]-\phi_{-i}[k]-\varepsilon[k])}{(\phi_i[k]+\phi_{-i}[k]+\varepsilon[k])^3}\right\}_{k\in\Z_T}\right)\,,
    \end{equation}
   \begin{equation}
        \frac{\partial^2\mathbf{J}_i}{\partial\mathbf{U}_{-i}\partial\mathbf{U}_i}=-\overline{\mathbf{W}}\:,\:\frac{\partial^2\mathbf{J}_i}{\partial\mathbf{U}_{-i}\partial\mathbf{\Phi}_i}=\mathbf{0}\:\text{and}\:\frac{\partial^2\mathbf{J}_i}{\partial\mathbf{\Phi}_{-i}\partial\mathbf{U}_i}=\mathbf{0}\,,
    \end{equation}
for $e=\mathbf{1}_2$, we have 
\begin{equation}
    \label{eq:21}
    \Gamma\defineas\left[\begin{array}{cc}
         2\mathbf{M}_{aa}& \mathbf{M}_s \\
         \mathbf{M}_s&2\mathbf{M}_{bb} 
    \end{array}\right]\,,
\end{equation}
where $\mathbf{M}_{aa}$, $\mathbf{M}_{bb}$ and $\mathbf{M}_s$ are given by
\begin{equation}
    \label{eq:22}
    \mathbf{M}_{aa}\defineas\left[\begin{array}{cc}
         \frac{\partial^2\mathbf{J}_a}{\partial\mathbf{\Phi}_a^2}& \mathbf{0} \\
         \mathbf{0}&\frac{\partial^2\mathbf{J}_a}{\partial\mathbf{U}_a^2} 
    \end{array}\right],\:\mathbf{M}_{bb}\defineas\left[\begin{array}{cc}
         \frac{\partial^2\mathbf{J}_b}{\partial\mathbf{\Phi}_b^2}& \mathbf{0} \\
         \mathbf{0}&\frac{\partial^2\mathbf{J}_b}{\partial\mathbf{U}_b^2} 
    \end{array}\right]\,,
\end{equation}
\begin{equation}
    \label{eq:23}
    \mathbf{M}_{s}\defineas\left[\begin{array}{cc}
         \frac{\partial^2\mathbf{J}_a}{\partial\mathbf{\Phi}_{b}\partial\mathbf{\Phi}_a}+\frac{\partial^2\mathbf{J}_b}{\partial\mathbf{\Phi}_{a}\partial\mathbf{\Phi}_b}& \mathbf{0} \\
         \mathbf{0}& \frac{\partial^2\mathbf{J}_a}{\partial\mathbf{U}_{b}\partial\mathbf{U}_a}+\frac{\partial^2\mathbf{J}_b}{\partial\mathbf{U}_{a}\partial\mathbf{U}_b}
    \end{array}\right]\,.
\end{equation}
It is clear that $\mathbf{M}_{aa}\prec0$. Now by utilizing the properties of the Shur complement~\cite[p.34]{Schur}, $\Gamma\prec0$ is equivalent to $\text{Sh}(2\mathbf{M}_{aa})\prec0$, where $\text{Sh}(2\mathbf{M}_{aa})\defineas 2\mathbf{M}_{bb}-\frac{1}{2}\mathbf{M}_{s}\mathbf{M}_{aa}^{-1}\mathbf{M}_{s}$ is the Schur complement of $2\mathbf{M}_{aa}$ in $\Gamma$. Therefore, by direct calculation we obtain $\text{Sh}(2\mathbf{M}_{aa})=\text{blkdiag}(\{\mathbf{S},-2\overline{\mathbf{W}}\})$,
where matrix $\mathbf{S}$ is given by
\begin{equation}
    \label{eq:25}
    \mathbf{S}=\text{diag}\biggl(\biggl\{\frac{-\beta[k](4\phi_a[k]+\varepsilon[k](4-\frac{\varepsilon[k]}{\varepsilon[k]+\phi_b[k]}))}{(\phi_a[k]+\phi_{b}[k]+\varepsilon[k])^3}\biggr\}_{k\in\Z_T}\biggr)\,.
\end{equation}
Since $\varepsilon[k]>0$ and $\phi_b[k]\geq0$, we have $\frac{\varepsilon[k]}{\varepsilon[k]+\phi_b[k]}\leq 1$ so $S\prec0$ and hence $\text{Sh}(2\mathbf{M}_{aa})\prec0$. This in return implies that $\Gamma \prec 0$ so the Nash equilibrium of $\mc G$ is unique.
\end{proof}}{

The proof is given in the extended version of our paper.}
\medskip

Having established the existence and uniqueness of the NE, in the following subsection we present a method based on theory of Variational inequalities~\cite{VIproblems} to compute it.

\subsection{Computing the Nash equilibrium}
Clearly, game $\mc G$ in~\eqref{eq:12} is equivalent to game $\overline{\mc G}$ given by
\begin{equation}\label{eq:26}
    \overline{\mc G}\defineas\biggl\{\min_{\mathbf{Z}_i\in\mc Z_i}-\textbf{J}_i(\mathbf{Z}_i,\mathbf{Z}_{-i}),\forall i\in\{a,b\}\biggr\}\,.
\end{equation}
Looking at the structure of $\overline{\mc G}$, we observe that the uncoupled nature of the constraint sets $\mathcal{Z}_i$, gives rise to a unique NE that coincides with the solution of the Variational Inequality problem $\text{VI}(\mc Z,F(\mathbf{Z}))$~\cite[Th.2]{Paccagnan2016a}, where $\mc Z=\mc Z_a\times\mc Z_b$ and $F(\mathbf{Z})=-g(\mathbf{Z},\mathbf{1}_2)$ given by
\begin{equation}\label{eq:27}
    F(\mathbf{Z})\defineas-\biggl[\bigl(\tfrac{\partial\mathbf{J}_a}{\partial\mathbf{\Phi}_a}\bigl)^T, \bigl(\tfrac{\partial\mathbf{J}_a}{\partial\mathbf{U}_a}\bigr)^T, \bigl(\tfrac{\partial\mathbf{J}_b}{\partial\mathbf{\Phi}_b}\bigr)^T, \bigl(\tfrac{\partial\mathbf{J}_b}{\partial\mathbf{U}_b}\bigr)^T \biggr]^T
\end{equation}
represents game $\overline{\mc G}$'s pseudogradient. Based on the structure of $F(\mathbf{Z})$, various iterative methods can be employed to compute the solution of $\text{VI}(\mc Z,F(\mathbf{Z}))$. Hence, we first establish the following characteristics of the pseudogradient $F(\mathbf{Z})$. 
\begin{lemma}\label{lema:1}
    Let game $\overline{\mc G}$ be defined by~\eqref{eq:19} and~\eqref{eq:26}, with the pseudogradient $F(\mathbf{Z})$ given by~\eqref{eq:27}. There exist $L,\mu>0$ such that for all $\mathbf{Z}_1,\mathbf{Z}_2\in\mc Z$, it holds that
    \begin{align}
        \norm{F(\mathbf{Z}_1)-F(\mathbf{Z}_2)}_2&\leq L\norm{\mathbf{Z}_1-\mathbf{Z}_2}_2,\\
        (F(\mathbf{Z}_1)-F(\mathbf{Z}_2))^T(\mathbf{Z}_1-&\mathbf{Z}_2)\geq\mu\norm{\mathbf{Z}_1-\mathbf{Z}_2}_2^2\,,
    \end{align}
    i.e, $F(\mathbf{Z})$ is $L$-Lipshitz continuous and $\mu$-strongly monotone. 
\end{lemma}

\iftoggle{full_version}{\begin{proof}
    To show that $F(\mathbf{Z})$ is Lipschitz continuous, we aim to invoke~\cite[P.2.3.2]{Lips}. For this, it suffices to show that every component of $F:\mc Z\rightarrow\R^{2T(m+1)}$ is Lipschitz continuous. Since $\mc Z_i$ in~\eqref{eq:19} is compact, it is clear that every component of the linear map $-\frac{\partial\mathbf{J}_i}{\partial\mathbf{U}_i}=\overline{\mathbf{W}}(2\mathbf{U}_i+\mathbf{U}_{-i})$ is Lipschitz continuous. Moreover, since $\varepsilon[k]>0$ and $\phi_i[k]+\phi_{-i}[k]\geq0$, every element of $-\frac{\partial\mathbf{J}_i}{\partial\mathbf{\Phi}_i}=\text{col}((p_k(\mathbf{Z}))_{k\in\Z_T})$, where 
    $$p_k:\mc Z\rightarrow\R\:\land\:p_k(\mathbf{Z})=\frac{-\beta[k](\phi_{-i}[k]+\varepsilon[k])}{(\phi_i[k]+\phi_{-i}[k]+\varepsilon[k])^2}\,,$$\normalsize
    is Lipshitz continuous since $\nabla_{\mathbf{Z}_i}p_k(\mathbf{Z})$ is continuous and the set $\left\{\norm{\nabla_{\mathbf{Z}_i}p_k(\mathbf{Z})}|\mathbf{Z}\in\mc Z\right\}$ is compact due to $\mc Z$ being compact. Strong monotonicity of $F(\mathbf{Z})$ follows directly from~\cite[P.2.3.2]{VIproblems} and the fact that $\Gamma=G(\mathbf{Z},\mathbf{1}_2)+G^{T}(\mathbf{Z},\mathbf{1}_2)$ in the proof of Theorem~\ref{th:1} is negative definite. 
\end{proof}}{

The proof is given in the extended version of our paper.}
\medskip

Based on Lemma~\ref{lema:1}, we can now outline in Theorem~\ref{th:2} a semi-decentralized iterative scheme that provably converges to the unique Nash equilibrium of the game defined in~\eqref{eq:12}.
\begin{theorem}\label{th:2}
    Let game $\mc G$ be defined as~\eqref{eq:12}, with the initial conditions $x_i^0$ such that constraint sets $\mc Z_i$ given by~\eqref{eq:19} satisfy Assumption~\ref{ass:2} for $i\in\{a,b\}$. Moreover, let $\mathbf{Z}_a^0\in\mc Z_a$ and  $\mathbf{Z}_b^0\in\mc Z_b$ be any feasible decision vectors for the two companies. Then, the iterative update scheme $\mathbf{Z}^{t+1}=\Pi_{\mc Z}\left[\mathbf{Z}^{t}-\gamma F(\mathbf{Z}^t)\right]$, that can be executed locally as,
    \begin{equation}\label{eq:30}
        \mathbf{Z}_i^{t+1}=\Pi_{\mc Z_i}\bigl[\mathbf{Z}_i^{t}+\gamma \nabla_{\mathbf{Z}_i}\mathbf{J}_i(\mathbf{Z}_i^{t},\mathbf{Z}_{-i}^{t})\bigr]\,,
    \end{equation}
    converges to the unique NE of $\mc G$ for $0<\gamma<\frac{2\mu}{L^2}$, where $L,\mu>0$ represent the Lipshitz and strong monotonicity constants of the pseudogradient operator $F(\mathbf{Z})$ defined in~\eqref{eq:27}. 
\end{theorem}
\iftoggle{full_version}{\begin{proof}
    Firstly, we note that the unique NE of $\mc G$ coincides with the unique NE of $\overline{\mc G}$ in~\eqref{eq:26} and that the constraint set $\mc Z$ is compact and convex as $\mc Z_i$ for $i\in\{a,b\}$ given by~\eqref{eq:19} is compact and convex. Now, for $0<\gamma<\frac{2\mu}{L^2}$, the mapping $\Pi_{\mc Z}\left[\mathbf{Z}-\gamma F(\mathbf{Z})\right]$ is a contraction from $\mc Z$ to $\mc Z$~\cite[T.12.1.2]{VIproblems} and converges to a fixed-point that corresponds to the unique NE of $\overline{\mc G}$~\cite[T.12.1.2]{VIproblems}, which completes the proof.
\end{proof}}{

The proof is given in the extended version of our paper.}
\subsection{Monitoring convergence and stopping criteria}
In essence, Theorem~\ref{th:2} ensures the existence of a sufficiently small fixed step size $\gamma>0$ that guarantees the convergence of~\eqref{eq:30}. However, neither Theorem~\ref{th:1} nor Lemma~\ref{lema:1} offer a method to explicitly determine the upper bound on~$\gamma$ as the Lipschitz and the strong monotonicity constants, $L$ and $\mu$, depend on the system dynamics and the initial states $x_i^0$ used to initialize~\eqref{eq:30}. To circumvent the need to estimate these constants through exhaustive grid-search across the space $\mathcal{Z}$ for each game instance, we propose a combination of a standard Armijo step-size rule with a termination criterion based on the Karush-Kuhn-Tucker (KKT) optimality conditions of the best-response optimization problems. 

Namely, for $\overline{\mathbf{Z}}\in\mc Z$ to be a NE of~\eqref{eq:12}, for each $i\in\{a,b\}$, the strategy $\overline{\mathbf{Z}}_i\in\mc Z_i$ has to solve the best-response problem 
\hspace{-1em}
\begin{mini!}
    {\mathbf{Z}_i}{-\textbf{g}_1(\mathbf{\Phi}_i,\overline{\mathbf{\Phi}}_{-i})-\textbf{g}_2(\mathbf{U}_i,\overline{\mathbf{U}}_{-i})}
    {\label{maxi:op1}}{}
    \addConstraint{\mathbf{L}_{\text{inq},i}\mathbf{Z}_i\leq\mathbf{r}_{\text{inq},i}\land \mathbf{L}_{\text{eq},i}\mathbf{Z}_i=\mathbf{r}_{\text{eq},i}.}{}{\label{eq:cc1}}
    \end{mini!}
Since this is a convex problem, and $\mc Z$ is compact and convex, characterizing the NE boils down to examining the KKT conditions associated with the Lagrangian 
\iftoggle{full_version}{\begin{multline}    \label{eq:31l}
    \mc L_i\defineas-\mathbf{J}_i(\mathbf{Z}_i,\overline{\mathbf{Z}}_{-i})+\mathbf{\lambda}_i^T(\mathbf{L}_{\text{inq},i}\mathbf{Z}_i-\mathbf{r}_{\text{inq},i})\\+\mathbf{\nu}^T_i(\mathbf{L}_{\text{eq},i}\mathbf{Z}_i-\mathbf{r}_{\text{eq},i})\,,
\end{multline}}{$\mc L_i\defineas-\mathbf{J}_i(\mathbf{Z}_i,\overline{\mathbf{Z}}_{-i})+\mathbf{\lambda}_i^T(\mathbf{L}_{\text{inq},i}\mathbf{Z}_i-\mathbf{r}_{\text{inq},i})+\mathbf{\nu}^T_i(\mathbf{L}_{\text{eq},i}\mathbf{Z}_i-\mathbf{r}_{\text{eq},i})$,}
with $\mathbf{\lambda}_i\in\R_+^{2mT}$ and $\mathbf{\nu}_i\in\R^{T}$ as dual variables associated with the inequality and equality constraints of each player $i\in\{a,b\}$.  With that in mind, we can construct the following optimality test for $\overline{\mathbf{Z}}\in\mathcal{Z}$.
\begin{lemma}[Optimality test]\label{lema:2}
    For players $i\in\{a,b\}$, let the best-response optimization problem related to game~\eqref{eq:26} be defined by~\eqref{maxi:op1}. Furthermore, for $\overline{\mathbf{Z}}\in\mc Z$, let the set $$\mc A_i(\overline{\mathbf{Z}}_i)\defineas\{j\in[1,2mT]\cap\N\:\:|\:\:\mathbf{L}_{\text{inq},i}^j\overline{\mathbf{Z}}_i=\mathbf{r}_{\text{inq},i}^j\},$$ represent the indices of active inequality constraints in~\eqref{eq:cc1}, with $\mathbf{L}_{\text{inq},i}^j$ being the $j$-th row of $\mathbf{L}_{\text{inq},i}$. Let $\delta_i^*(\overline{\mathbf{Z}})\in\R_+$ be
    \begin{equation}\label{eq:31}
        \begin{array}{c}
        \delta_i^*(\overline{\mathbf{Z}})=\displaystyle\min _{\lambda_i,\nu_i}  \norm{-\nabla_{\mathbf{Z}_i}\mathbf{J}_i(\overline{\mathbf{Z}}_i,\overline{\mathbf{Z}}_{-i})+\mathbf{L}_{\text{inq},i}^T\lambda_i+\mathbf{L}_{\text{eq},i}^T\nu_i}^2_{2} \\
        \text {subject to } \lambda_i^j\geq 0 \text{ for }j\in\mc A_i(\overline{\mathbf{Z}}_i)\text{ and }\lambda_i^j=0 \text{ otherwise}. 
        \end{array}
    \end{equation}
    Then, $\overline{\mathbf{Z}}\in\mc Z$ is the solution of the best-response optimization problem~\eqref{maxi:op1} if and only if $\delta_a^*(\overline{\mathbf{Z}})=\delta_b^*(\overline{\mathbf{Z}})=0$.
\end{lemma}
\iftoggle{full_version}{\begin{proof}
    For each player $i\in\{a,b\}$, the condition $\delta_i^*(\overline{\mathbf{Z}})=0$ coincides with the existence of feasible $\lambda_i$ and $\nu_i$ such that $-\nabla_{\mathbf{Z}_i}\mathbf{J}_i(\overline{\mathbf{Z}}_i,\overline{\mathbf{Z}}_{-i})+\mathbf{L}_{\text{inq},i}^T\lambda_i+\mathbf{L}_{\text{eq},i}^T\nu_i=\mathbf{0}$. This equation precisely represents the stationarity condition of~\eqref{maxi:op1} for $\mathbf{Z}_i$. Conversely, the feasibility of $\lambda_i$, as dictated by the constraints of~\eqref{eq:31}, corresponds exactly to the complementarity slackness constraint of~\eqref{maxi:op1}. Finally, since problem~\eqref{maxi:op1} is convex, the optimality of $\mathbf{Z_i}$ is equivalent to finding a solution of the KKT system, thereby completing the proof.
\end{proof}}{

The proof is given in the extended version of our paper.}
\medskip

We can now combine Lemma~\ref{lema:2} with an Armijo-like approach, to design a complete iterative procedure independent of constants $L$ and $\mu$ that are specific to the game. 

Namely, for a priori chosen constants $\overline{\gamma}>0$ and $\eta\in(0,1)$, the complete procedure is outlined in Algorithm~\ref{al:1}. 
\begin{algorithm}[tbp]
    \caption{Computing the Nash equilibrium}\label{al:1}
    \begin{algorithmic}[1]
    \State \textbf{Input:}  $A_i$, $B_i$, $T$, $\varepsilon$, $\beta$, $\overline{\mathbf{W}}$, $x_i^0$, $\overline{\gamma}$, $\eta$, $\text{tol}$, $\text{maxiter}$
    \State \textbf{Output:} $\mathbf{Z}_a,\mathbf{Z}_b$
    \State $\mc Z_i=\text{CreateConstraints}(T,A_i,B_i,x_i^0)$;\Comment{For $i\in\{a,b\}$}
    \State $\mathbf{Z}_i^0=\text{Initialize}(\mc Z_i,x_i^0)$;
    \State $l=0;\quad\mathbf{Z}\leftarrow\text{col}((\mathbf{Z}_i^0)_{i\in\{a,b\}})$;
    \While {$\sum_{i\in\{a,b\}}\delta_i^*(\mathbf{Z})>\text{tol}$}\Comment{In parallel for $i$}
        \State $\gamma=\eta^l\overline{\gamma};\quad t=0$;
        \State $\mathbf{Z}_i^t\leftarrow\mathbf{Z}_i^0$;
        \State $\mathbf{Z}_i^{t+1}=\Pi_{\mc Z_i}\left[\mathbf{Z}_i^{t}+\gamma \nabla_{\mathbf{Z}_i}\mathbf{J}_i(\mathbf{Z}_i^{t},\mathbf{Z}_{-i}^{t})\right]$;
        \While{$\norm{\mathbf{Z}_i^{t+1}-\mathbf{Z}_i^t}>\text{tol}\textbf{ and }t<\text{maxiter}$}
            \State $\mathbf{Z}_i^t\leftarrow\mathbf{Z}_i^{t+1}$;
            \State $\mathbf{Z}_i^{t+1}=\Pi_{\mc Z_i}\left[\mathbf{Z}_i^{t}+\gamma \nabla_{\mathbf{Z}_i}\mathbf{J}_i(\mathbf{Z}_i^{t},\mathbf{Z}_{-i}^{t})\right]$;  
            \State $t\leftarrow t+1$;
        \EndWhile
        \State $\mathbf{Z}\leftarrow\text{col}((\mathbf{Z}_i^{t+1})_{i\in\{a,b\}})$
        \State $l\leftarrow l+1$  
     \EndWhile
    \end{algorithmic}
\end{algorithm}
For a particular step size $\gamma=\eta^l\overline{\gamma}$ with $l\in\Z_+$, the inner loop is used to monitor convergence of the iterates $\mathbf{Z}_i^t$ obtained using scheme~\eqref{eq:30}. After convergence or timeout have been detected, Lemma~\ref{lema:2} is then used in the condition of the outer loop to check optimality of the obtained solution. If conditions of the optimality test are satisfied, the obtained solution corresponds to the NE of game~\eqref{eq:12}. If not, the step size is reduced and the procedure is repeated. Finally, it is important to note that the outer loop will always terminate as Theorem~\ref{th:2} guarantees the existence of a sufficiently small step size for which~\eqref{eq:30} converges to the NE of~\eqref{eq:12}.

\subsection{Receding-horizon implementation}
In essence, we can deploy Algorithm~\ref{al:1} in two ways:
\begin{enumerate}
    \item In an open-loop manner, i.e., the exogenous parameters can be estimated for the whole time frame so the planning horizon can be chosen as $T=T_{\text{total}}$.
    \item In a closed-loop, MPC-like, receding-horizon fashion, i.e., the planning horizon is $T<T_{\text{total}}$, for example due to lack of accurate predictions, and, at each time step $k$, the companies apply only the first element of the predicted control trajectory. 
\end{enumerate}
When applied in the open-loop manner, Algorithm~\ref{al:1} is executed once to compute $T_{\text{total}}$ control inputs. On the other hand, running it in a receding-horizon fashion requires the complete procedure to be repeated $n_{\text{total}}=T_{\text{total}}-T+1$ times. For time intervals $0\leq k\leq n_{\text{total}}-1$, the planning is executed with horizon $T$ and only the first element of the predicted control trajectory is applied, i.e., the first $T_{\text{total}}-T$ control inputs yield a closed-loop system of the form $x_i[k+1]=A_ix_i[k]+B_i\kappa_i(x_i[k],T)$, 
where $\kappa_i(x_i[k],T)$ extracts $u_i[0]$ from the output $\mathbf{Z}_i$ of Algortihm~\ref{al:1} when initialized with $x_i^0=x_i[k]$. Conversely, the control inputs for the last $T$ time intervals are applied in the open-loop fashion and are calculated in the final execution step, i.e., for $k=n_{\text{total}}$.

In the subsequent section, we introduce a numerical case study wherein the proposed method will be utilized both as an open-loop control strategy and in a receding-horizon manner to formulate feedback policies inspired by MPC. 
\iftoggle{full_version}{\begin{figure}[!t]
    \centering
    \begin{adjustbox}{ max width=0.48\textwidth}
    \usepgfplotslibrary{groupplots}

\begin{tikzpicture}

\definecolor{color0}{rgb}{1,0.498039215686275,0.0549019607843137}
\definecolor{color1}{rgb}{0.580392156862745,0.403921568627451,0.741176470588235}

\begin{groupplot}[group style={group size=1 by 3}]
\nextgroupplot[
legend cell align={right},
legend style={fill opacity=0.8, draw opacity=1, text opacity=1, draw=white!90!black,at={(1,1)}, font=\small},
tick align=outside,
tick pos=left,
x grid style={white!90!black},
xmajorgrids,
xmin=-0.1, xmax=9.1,
xtick style={color=black},
y grid style={white!90!black},
ylabel=\textcolor{black}{$\beta[k]$},
ymin=-2750, ymax=167750,
ytick style={color=black},
height=2.7cm,
width=8cm
]
\addplot [semithick, black, const plot mark left, dashed,  mark size=3, mark options={solid}]
table {%
0 5000
1 5000
2 80000
3 160000
4 140000
5 100000
6 20000
7 5000
8 5000
9 5000
};

\nextgroupplot[
legend cell align={right},
legend style={fill opacity=0.8, draw opacity=1, text opacity=1, draw=white!90!black,at={(1,1)}, font=\small},
tick align=outside,
tick pos=left,
x grid style={white!90!black},
xmajorgrids,
xmin=-0.1, xmax=9.1,
xtick style={color=black},
y grid style={white!90!black},
ylabel=\textcolor{black}{$q[k]$},
ylabel style={xshift=-4pt},
ymin=0.0, ymax=1.6,
ytick style={color=black},
height=2.7cm,
width=8cm
]
\addplot [semithick, black, const plot mark left, dashed,  mark size=3, mark options={solid}]
table {%
0 1
1 1
2 0.1
3 0.1
4 0.1
5 0.5
6 1.5
7 1.5
8 1.5
9 1.5
};

\nextgroupplot[
legend cell align={right},
legend style={fill opacity=0.8, draw opacity=1, text opacity=1, draw=white!90!black,at={(1,1)}, font=\small},
tick align=outside,
tick pos=left,
x grid style={white!90!black},
xlabel={Time interval $k$},
xmajorgrids,
xmin=-0.1, xmax=9.1,
xtick style={color=black},
y grid style={white!90!black},
ylabel=\textcolor{black}{$\varepsilon[k]$},
ylabel style={xshift=-4pt},
ymin=8, ymax=52,
ytick style={color=black},
height=2.7cm,
width=8cm,
ytick={10, 30, 50},
]
\addplot [semithick, black, const plot mark left, dashed,  mark size=3, mark options={solid}]
table {%
0 10
1 20
2 30
3 50
4 50
5 40
6 20
7 10
8 10
9 10
};

\end{groupplot}

\end{tikzpicture}
    \end{adjustbox}
    \caption{Time profiles of $\beta[k]$, $q[k]$ and $\varepsilon[k]$, determining the expected ride-hailing revenue, charging costs and customer abandonments.}
    \label{fig:setup}
\end{figure}}{
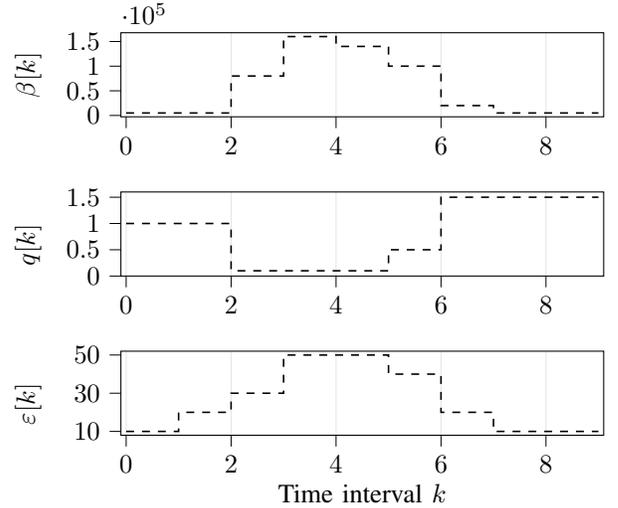
\begin{figure}[!t]
    \centering
    \begin{adjustbox}{ max width=0.48\textwidth}
    \usepgfplotslibrary{groupplots}

\begin{tikzpicture}

\definecolor{color0}{rgb}{1,0.498039215686275,0.0549019607843137}
\definecolor{color1}{rgb}{0.580392156862745,0.403921568627451,0.741176470588235}

\begin{groupplot}[group style={group size=2 by 1}]
\nextgroupplot[
legend cell align={right},
legend style={fill opacity=0.8, draw opacity=1, text opacity=1, draw=white!90!black,at={(1,1)}, font=\small},
tick align=outside,
tick pos=left,
x grid style={white!90!black},
xlabel={Time interval $k$},
xmajorgrids,
xmin=-0.1, xmax=9.1,
xtick style={color=black},
y grid style={white!90!black},
ymin=0.0, ymax=1.6,
ytick style={color=black},
height=3.6cm,
width=5cm
]
\addplot [semithick, black, const plot mark left, dashed,  mark size=3, mark options={solid}]
table {%
0 0.05
1 0.05
2 0.80000
3 1.60000
4 1.40000
5 1.00000
6 0.20000
7 0.05000
8 0.05000
9 0.05000
};
\addlegendentry{$\beta[k]$}

\addplot [semithick, black, const plot mark left, dotted,  mark size=3, mark options={solid}]
table {%
0 1
1 1
2 0.1
3 0.1
4 0.1
5 0.5
6 1.5
7 1.5
8 1.5
9 1.5
};
\addlegendentry{$q[k]$}

\nextgroupplot[
legend cell align={right},
legend style={fill opacity=0.8, draw opacity=1, text opacity=1, draw=white!90!black,at={(1,1)}, font=\small},
tick align=outside,
tick pos=left,
x grid style={white!90!black},
xlabel={Time interval $k$},
xmajorgrids,
xmin=-0.1, xmax=9.1,
xtick style={color=black},
y grid style={white!90!black},
ylabel style={xshift=-4pt},
ymin=8, ymax=52,
ytick style={color=black},
height=3.6cm,
width=5cm,
ytick={10, 30, 50},
]
\addplot [semithick, black, const plot mark left, dashed,  mark size=3, mark options={solid}]
table {%
0 10
1 20
2 30
3 50
4 50
5 40
6 20
7 10
8 10
9 10
};
\addlegendentry{$\varepsilon[k]$}
\end{groupplot}
\end{tikzpicture}
    \end{adjustbox}
    \vspace{-1.5em}
    \caption{Time profiles of $\beta[k]$, $q[k]$ and $\varepsilon[k]$, determining the expected ride-hailing revenue, charging costs and customer abandonments.}
    \label{fig:setup}
\vspace{-1.5em}
\end{figure}

}
\section{Numerical Example}\label{sec:example}

To demonstrate the effectiveness of the proposed method, we construct a scenario spanning $T_{\text{total}}=9$ time intervals, depicting varying levels of ride-hailing demand, electricity costs, and anticipated customer abandonments, as illustrated in Figure~\ref{fig:setup}. For simplicity, we assume that each vehicle's battery level can be categorized as green $(j=2)$, yellow $(j=1)$, or red $(j=0)$. We further assume that the coefficients in model~\eqref{eq:1} satisfy $\alpha_i^j=0$ for all $i\in\{a,b\}$ and $j\in\mathbb{Z}_3$, and that the charging cost coefficients are identical across all battery level categories, i.e., $q^j[k]=q[k]$ for all $j\in\mathbb{Z}_3$. Regarding the initial state of the electric fleets, we assume that $x_a^0=[400,50,10]^T$ and $x_b^0=[800,50,10]^T$. As outlined in Section~\ref{sec:NE}, we test Algorithm~\ref{al:1} in an open-loop manner, i.e., with $T=T_{\text{total}}$, and in a closed-loop manner, i.e., with $T<T_{\text{total}}$.
\iftoggle{full_version}{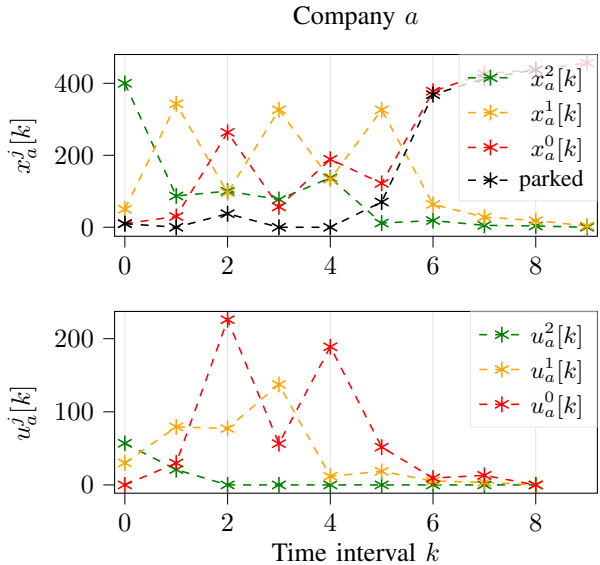
\begin{figure}[!t]
    \centering
    \begin{adjustbox}{ max width=0.48\textwidth}
    \usepgfplotslibrary{groupplots}

\begin{tikzpicture}

\definecolor{color0}{rgb}{1,0.647058823529412,0}

\begin{groupplot}[group style={group size=1 by 2}]
\nextgroupplot[
legend cell align={right},
legend style={fill opacity=0.8, draw opacity=1, text opacity=1, draw=white!90!black,at={(1,1)}, font=\small},
tick align=outside,
tick pos=left,
x grid style={white!90!black},
xmajorgrids,
xmin=-0.2, xmax=9.2,
xtick style={color=black},
y grid style={white!90!black},
ylabel=\textcolor{black}{$x_a^j[k]$},
ymin=-25, ymax=480,
ytick style={color=black},
height=4cm,
width=8cm,
title={Company $a$}
]
\addplot [semithick, green!50.1960784313725!black, dashed, mark=asterisk, mark size=3, mark options={solid}]
table {%
0 400
1 87.0000235320865
2 100.37308523481
3 77.1716119544859
4 137.460770597194
5 11.7195718367531
6 18.638867406831
7 5.32702426642696
8 3.59936213438774
9 6.73718358834781e-23
};
\addlegendentry{$x_a^2[k]$}
\addplot [semithick, color0, dashed, mark=asterisk, mark size=3, mark options={solid}]
table {%
0 50
1 343.066693359526
2 95.9706496402796
3 326.358496507631
4 133.641503492369
5 326.358496507631
6 63.4000890085323
7 28.1161553421785
8 18.4036743361622
9 3.59936213438774
};
\addlegendentry{$x_a^1[k]$}
\addplot [semithick, red, dashed, mark=asterisk, mark size=3, mark options={solid}]
table {%
0 10
1 29.9332831083872
2 263.65626512491
3 56.4698915378827
4 188.897725910437
5 121.921931655616
6 377.961043584637
7 426.556820391395
8 437.99696352945
9 456.400637865612
};
\addlegendentry{$x_a^0[k]$}
\addplot [semithick, black, dashed, mark=asterisk, mark size=3, mark options={solid}]
table {%
0 10
1 -3.5527136788005e-15
2 37.6708538520891
3 -7.105427357601e-15
4 -2.8421709430404e-14
5 70.2414144838364
6 368.483755649289
7 413.480170321659
8 437.99696352945
};
\addlegendentry{parked}

\nextgroupplot[
legend cell align={right},
legend style={fill opacity=0.8, draw opacity=1, text opacity=1, draw=white!90!black,at={(1,1)}, font=\small},
tick align=outside,
tick pos=left,
x grid style={white!90!black},
xmajorgrids,
xmin=-0.2, xmax=9.2,
xtick style={color=black},
xlabel={Time interval $k$},
y grid style={white!90!black},
ylabel=\textcolor{black}{$u_a^j[k]$},
ylabel style={xshift=-4pt},
ymin=-11.2992705636411, ymax=237.284681836462,
ytick style={color=black},
height=4cm,
width=8cm
]
\addplot [semithick, green!50.1960784313725!black, dashed, mark=asterisk, mark size=3, mark options={solid}]
table {%
0 56.9333066404737
1 20.9626570001941
2 4.25416804535969e-22
3 -1.02061899181261e-23
4 6.41725144434117e-22
5 3.93804823080058e-22
6 1.03058719141085e-22
7 8.68285709604338e-23
8 3.36859179417391e-23
};
\addlegendentry{$u_a^2[k]$}
\addplot [semithick, color0, dashed, mark=asterisk, mark size=3, mark options={solid}]
table {%
0 30.0667168916128
1 79.4104282346163
2 77.1716119544859
3 137.460770597194
4 11.7195718367531
5 18.638867406831
6 5.32702426642696
7 3.59936213438774
8 3.36859179417391e-23
};
\addlegendentry{$u_a^1[k]$}
\addplot [semithick, red, dashed, mark=asterisk, mark size=3, mark options={solid}]
table {%
0 -1.09928652902678e-22
1 29.9332831083872
2 225.985411272821
3 56.4698915378828
4 188.897725910437
5 51.6805171717792
6 9.47728793534751
7 13.0766500697353
8 0
};
\addlegendentry{$u_a^0[k]$}

\end{groupplot}

\end{tikzpicture}
    \end{adjustbox}
    \caption{\textbf{Open loop:} Evolution of the company $a$'s state and applied control input during time frame $T_{\text{total}}$.}
    \label{fig:t9a}
\end{figure}
\begin{figure}[!t]
    \centering
    \begin{adjustbox}{ max width=0.48\textwidth}
    \usepgfplotslibrary{groupplots}

\begin{tikzpicture}

\definecolor{color0}{rgb}{1,0.647058823529412,0}

\begin{groupplot}[group style={group size=1 by 2}]
\nextgroupplot[
legend cell align={right},
legend style={fill opacity=0.8, draw opacity=1, text opacity=1, draw=white!90!black,at={(1,1)}, font=\small},
tick align=outside,
tick pos=left,
x grid style={white!90!black},
xmajorgrids,
xmin=-0.2, xmax=9.2,
xtick style={color=black},
y grid style={white!90!black},
ylabel=\textcolor{black}{$x_b^j[k]$},
ymin=-25, ymax=902,
ytick style={color=black},
height=4cm,
width=8cm,
title={Company $b$}
]
\addplot [semithick, green!50.1960784313725!black, dashed, mark=asterisk, mark size=3, mark options={solid}]
table {%
0 800
1 77.3861316723912
2 117.046572463972
3 59.5892577008259
4 129.936326155594
5 100.806339348657
6 12.8725846273189
7 14.2343505683492
8 1.47615953030073
9 1.31452611191447e-22
};
\addlegendentry{$x_b^2[k]$}
\addplot [semithick, color0, dashed, mark=asterisk, mark size=3, mark options={solid}]
table {%
0 50
1 752.612314634024
2 72.5348050948618
3 465.201209127164
4 394.798790872836
5 331.620055003351
6 117.613836864841
7 19.0739675557217
8 23.8978402158511
9 1.47615953030073
};
\addlegendentry{$x_b^1[k]$}
\addplot [semithick, red, dashed, mark=asterisk, mark size=3, mark options={solid}]
table {%
0 10
1 30.0015536935851
2 670.418622441167
3 335.20953317201
4 335.26488297157
5 427.573605647992
6 729.51357850784
7 826.691681875929
8 834.626000253848
9 858.523840469699
};
\addlegendentry{$x_b^0[k]$}
\addplot [semithick, black, dashed, mark=asterisk, mark size=3, mark options={solid}]
table {%
0 10
1 -7.105427357601e-15
2 322.263985777974
3 1.70530256582424e-13
4 133.581154123813
5 410.766108131808
6 721.326248390639
7 817.028192228427
8 834.626000253848
};
\addlegendentry{parked}

\nextgroupplot[
legend cell align={right},
legend style={fill opacity=0.8, draw opacity=1, text opacity=1, draw=white!90!black,at={(1,1)}, font=\small},
tick align=outside,
tick pos=left,
x grid style={white!90!black},
xmajorgrids,
xmin=-0.2, xmax=9.2,
xtick style={color=black},
xlabel={Time interval $k$},
y grid style={white!90!black},
ylabel=\textcolor{black}{$u_b^j[k]$},
ylabel style={xshift=-4pt},
ymin=-11.2992705636411, ymax=366,
ytick style={color=black},
height=4cm,
width=8cm
]
\addplot [semithick, green!50.1960784313725!black, dashed, mark=asterisk, mark size=3, mark options={solid}]
table {%
0 47.3876853659763
1 34.8528802711145
2 -7.94715897541771e-23
3 1.11619825974274e-21
4 7.10238265329059e-22
5 1.97623027196524e-22
6 1.98594718879837
7 3.64841964306169e-23
8 6.57263055957233e-23
};
\addlegendentry{$u_b^2[k]$}
\addplot [semithick, color0, dashed, mark=asterisk, mark size=3, mark options={solid}]
table {%
0 29.9984463064149
1 82.1936921928571
2 59.5892577008259
3 129.936326155594
4 100.806339348657
5 12.8725846273189
6 12.2484033795508
7 1.47615953030073
8 6.57263055957233e-23
};
\addlegendentry{$u_b^1[k]$}
\addplot [semithick, red, dashed, mark=asterisk, mark size=3, mark options={solid}]
table {%
0 4.22626454186321e-23
1 30.0015536935851
2 348.154636663192
3 335.20953317201
4 201.683728847757
5 16.807497516184
6 8.18733011720116
7 9.66348964750188
8 0
};
\addlegendentry{$u_b^0[k]$}

\end{groupplot}

\end{tikzpicture}
    \end{adjustbox}
    \caption{\textbf{Open loop:} Evolution of the company $b$'s state and applied control input during time frame $T_{\text{total}}$.}
    \label{fig:t9b}
\end{figure}}{
\begin{figure}[!t]
    \centering
    \begin{adjustbox}{ max width=0.48\textwidth}
    \usepgfplotslibrary{groupplots}

\begin{tikzpicture}

\definecolor{color0}{rgb}{1,0.647058823529412,0}

\begin{groupplot}[group style={group size=1 by 2}]

\nextgroupplot[
legend cell align={right},
legend style={fill opacity=0.8, draw opacity=1, text opacity=1, draw=white!90!black,at={(1,1)}, font=\small},
tick align=outside,
tick pos=left,
x grid style={white!90!black},
xmajorgrids,
xmin=-0.2, xmax=9.2,
xtick style={color=black},
y grid style={white!90!black},
ylabel=\textcolor{black}{$u_a^j[k]$},
ylabel style={xshift=-4pt},
ymin=-11.2992705636411, ymax=237.284681836462,
ytick style={color=black},
height=3.2cm,
width=8cm
]
\addplot [semithick, green!50.1960784313725!black, dashed, mark=asterisk, mark size=3, mark options={solid}]
table {%
0 56.9333066404737
1 20.9626570001941
2 4.25416804535969e-22
3 -1.02061899181261e-23
4 6.41725144434117e-22
5 3.93804823080058e-22
6 1.03058719141085e-22
7 8.68285709604338e-23
8 3.36859179417391e-23
};
\addlegendentry{$u_a^2[k]$}
\addplot [semithick, color0, dashed, mark=asterisk, mark size=3, mark options={solid}]
table {%
0 30.0667168916128
1 79.4104282346163
2 77.1716119544859
3 137.460770597194
4 11.7195718367531
5 18.638867406831
6 5.32702426642696
7 3.59936213438774
8 3.36859179417391e-23
};
\addlegendentry{$u_a^1[k]$}
\addplot [semithick, red, dashed, mark=asterisk, mark size=3, mark options={solid}]
table {%
0 -1.09928652902678e-22
1 29.9332831083872
2 225.985411272821
3 56.4698915378828
4 188.897725910437
5 51.6805171717792
6 9.47728793534751
7 13.0766500697353
8 0
};
\addlegendentry{$u_a^0[k]$}

\nextgroupplot[
legend cell align={right},
legend style={fill opacity=0.8, draw opacity=1, text opacity=1, draw=white!90!black,at={(1,1)}, font=\small},
tick align=outside,
tick pos=left,
x grid style={white!90!black},
xmajorgrids,
xmin=-0.2, xmax=9.2,
xtick style={color=black},
xlabel={Time interval $k$},
y grid style={white!90!black},
ylabel=\textcolor{black}{$u_b^j[k]$},
ylabel style={xshift=-4pt},
ymin=-11.2992705636411, ymax=366,
ytick style={color=black},
height=3.2cm,
width=8cm
]
\addplot [semithick, green!50.1960784313725!black, dashed, mark=asterisk, mark size=3, mark options={solid}]
table {%
0 47.3876853659763
1 34.8528802711145
2 -7.94715897541771e-23
3 1.11619825974274e-21
4 7.10238265329059e-22
5 1.97623027196524e-22
6 1.98594718879837
7 3.64841964306169e-23
8 6.57263055957233e-23
};
\addlegendentry{$u_b^2[k]$}
\addplot [semithick, color0, dashed, mark=asterisk, mark size=3, mark options={solid}]
table {%
0 29.9984463064149
1 82.1936921928571
2 59.5892577008259
3 129.936326155594
4 100.806339348657
5 12.8725846273189
6 12.2484033795508
7 1.47615953030073
8 6.57263055957233e-23
};
\addlegendentry{$u_b^1[k]$}
\addplot [semithick, red, dashed, mark=asterisk, mark size=3, mark options={solid}]
table {%
0 4.22626454186321e-23
1 30.0015536935851
2 348.154636663192
3 335.20953317201
4 201.683728847757
5 16.807497516184
6 8.18733011720116
7 9.66348964750188
8 0
};
\addlegendentry{$u_b^0[k]$}

\end{groupplot}

\end{tikzpicture}
    \end{adjustbox}
    \vspace{-0.5em}
    \caption{Open loop: Evolution of the applied control inputs during $T_{\text{total}}$.}
    \label{fig:t9c}
\end{figure}}
In both cases we set the tolerance to 0.01 and initialize the step size with $\overline{\gamma}=0.05$ and $\mu=0.5$. 

\iftoggle{full_version}{ Figures~\ref{fig:t9a} and~\ref{fig:t9b} depict the evolution of the system state and the optimal inputs when employing the proposed method in an open-loop fashion, i.e., with a planning horizon of $T=9$.}{Figure~\ref{fig:t9c} depicts the evolution of the optimal inputs when employing the proposed method in an open-loop fashion, i.e., with a planning horizon of $T=9$.} It can be observed that both companies predominantly utilize intervals characterized by lower electricity prices to recharge their vehicles, aiming to strike a balance between maintaining a high number of operational vehicles during peak demand periods and ensuring that vehicle batteries remain sufficiently charged to prevent forcing them to park. Furthermore, towards the end of the analyzed time period, both companies opt not to charge but rather to park their vehicles.\iftoggle{full_version}{ This is anticipated since there are no parking costs associated, and there is no longer a need to proactively charge vehicles as the time frame concludes. Similar observations are referred to as the 'end-of-day' effects in the literature~\cite{6994293, 9354436, ESTRELLA2019126}. Looking at Figure~\ref{fig:profit}, we can observe that during the initial two time intervals, the companies incur negative profits. This is due to their proactive preparation for the demand peak, which leads to high charging costs outweighing the low revenue generated from providing ride-hailing services.}{ This is anticipated since there are no parking costs associated, and there is no longer a need to proactively charge vehicles.}

To examine how the performance of the proposed method varies with different planning horizons $T$ in a receding-horizon approach, we also execute the procedure for $T=3$ and $T=6$ and apply it in the receding horizon fashion.
\begin{table}[!t]
\begin{center}
 \renewcommand{\arraystretch}{1.3}
 \caption{Attained profits for different planning horizons}\vspace{1ex}
 \label{tab:res}
 \begin{tabular}{c|c|c|c}
 Horizon  & Company $a$ & Company $b$ & Lost profit \\ 
 \hline\hline
 $T=3$ &  88315 & 75275    & 250363    \\
 $T=6$ &  143859 & 207973  & 42622   \\
 $T=9$ &  144999 & 211129  & 38115   \\
\hline 
\end{tabular}
\end{center}
\vspace{-2em}
\end{table}
Table~\ref{tab:res} summarizes the total attained profits of the companies for different horizons\iftoggle{full_version}{ whereas Figures~\ref{fig:profit6} and~\ref{fig:profit3} illustrate the attained profits for every time interval $0\leq k\leq T_{\text{total}}-1$}{}. As expected, extending the planning horizon enables companies to proactively anticipate fluctuations in demand and charging costs, facilitating better charging strategies and ultimately leading to higher overall profits.\iftoggle{full_version}{ This trend is supported by numerical observations in Table~\ref{tab:res}, where both companies demonstrate an increase in total profit as the planning horizon expands.}{} Moreover, with longer planning horizons, companies tend to better plan their participation, leading to smaller amounts of profit forfeited due to customer abandonment. \iftoggle{full_version}{ It is interesting, however, to observe that for $T=3$, company $a$ attains higher total profit than company $b$ over the analyzed time period, even though it operates a smaller fleet.

Before concluding the numerical case study, it is worth noting that operating the proposed method in an open-loop manner cannot respond to sudden disturbances in the stochastic ride-hailing demand. Therefore, it would be beneficial to analyze the closed-loop stability of the system in order to help enhance the robustness of the charging planning strategies. This is beyond the scope of this paper and is considered an interesting direction for future research.}{}  
\section{Conclusions}\label{sec:conclusion}
In this work, we presented a dynamic, game-theoretic framework for optimizing the charging schedule of ride-hailing companies over a predefined time period for which the estimates of the exogenous parameters are available. The proposed setup takes into account the varying time profiles of the expected ride-hailing demand, charging prices, and expected customer abandonments, all incorporated in a modified Tullock-based profit function. For this game that describes interactions between two ride-hailing companies over a fixed planning horizon, we were able to show that there exists a unique NE that can be computed using a semi-decentralized iterative method, requiring little information exchange. The proposed method is tested as an open-loop planning module and in a receding-horizon manner, showing better charging management for longer planning horizons. 

\iftoggle{full_version}{ In the future, we aim to examine the closed-loop stability in an attempt to also provide certain robustness guarantees. Moreover, it would be beneficial to extend the analysis to cases with more than two players and setups where companies have different estimates of the demand and charging price-related parameters. We also plan to investigate how electricity market models can be integrated with the proposed model. }{In the future, we aim to examine the closed-loop stability in an attempt to also provide robustness guarantees. Moreover, we hope to extend the analysis to cases with more than two players and setups where companies have different estimates of the relevant parameters. We also plan to investigate how electricity market models can be integrated with the proposed model. }    
\iftoggle{full_version}{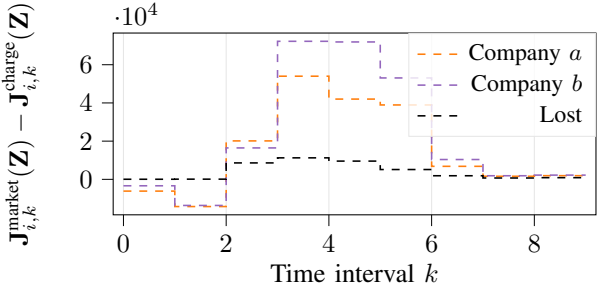
\begin{figure}[!t]
    \centering
    \begin{adjustbox}{ max width=0.48\textwidth}
\begin{tikzpicture}

\definecolor{color0}{rgb}{1,0.498039215686275,0.0549019607843137}
\definecolor{color1}{rgb}{0.580392156862745,0.403921568627451,0.741176470588235}

\begin{axis}[
legend cell align={right},
legend style={fill opacity=0.8, draw opacity=1, text opacity=1, draw=white!90!black,at={(1,1)}, font=\small},
tick align=outside,
tick pos=left,
x grid style={white!90!black},
xlabel={Time interval $k$},
xmajorgrids,
xmin=-0.2, xmax=9.2,
xtick style={color=black},
y grid style={white!90!black},
ylabel=\textcolor{black}{$\mathbf{J}_{i,k}^{\text{market}}(\mathbf{Z})-\mathbf{J}_{i,k}^{\text{charge}}(\mathbf{Z})$},
ymin=-18571.2316110751, ymax=76587.0800394689,
ytick style={color=black},
height=4cm,
width=8cm,
]
\addplot [semithick, color0, const plot mark left, dashed,  mark size=3, mark options={solid}]
table {%
0 -6160.99790748319
1 -14245.8538087776
2 20120.9806720169
3 53993.9306641415
4 42010.1211763044
5 38943.2669319759
6 6812.65973218686
7 1608.40613508384
8 1917.40789707653
9 1917.40789707653
};
\addlegendentry{Company $a$}

\addplot [semithick, color1, const plot mark left, dashed,  mark size=3, mark options={solid}]
table {%
0 -3373.3400196741
1 -13671.7015534788
2 16447.8967463725
3 72261.7022371714
4 71922.5444654276
5 53080.1089247172
6 10371.7710104403
7 1879.69774969478
8 2211.16333460648
9 2211.16333460648
};
\addlegendentry{Company $b$}

\addplot [semithick, black, const plot mark left, dashed,  mark size=3, mark options={solid}]
table {%
0 43.6447239417959
1 94.1047597552475
2 8597.08957032616
3 11252.9691893979
4 9545.73900935459
5 5134.83426070728
6 1878.25169619563
7 697.583803429098
8 871.428768316995
9 871.428768316995
};
\addlegendentry{Lost}

\end{axis}

\end{tikzpicture}
    \end{adjustbox}
    \caption{\textbf{Open loop:} Attained profits for $T=9$.}
    \label{fig:profit}
\end{figure}
\begin{figure}[!t]
    \centering
    \begin{adjustbox}{ max width=0.48\textwidth}
\begin{tikzpicture}

\definecolor{color0}{rgb}{1,0.498039215686275,0.0549019607843137}
\definecolor{color1}{rgb}{0.580392156862745,0.403921568627451,0.741176470588235}

\begin{axis}[
legend cell align={right},
legend style={fill opacity=0.8, draw opacity=1, text opacity=1, draw=white!90!black,at={(1,1)}, font=\small},
tick align=outside,
tick pos=left,
x grid style={white!90!black},
xlabel={Time interval $k$},
xmajorgrids,
xmin=-0.2, xmax=9.2,
xtick style={color=black},
y grid style={white!90!black},
ylabel=\textcolor{black}{$\mathbf{J}_{i,k}^{\text{market}}(\mathbf{Z})-\mathbf{J}_{i,k}^{\text{charge}}(\mathbf{Z})$},
ymin=-18571.2316110751, ymax=76587.0800394689,
ytick style={color=black},
height=4cm,
width=8cm,
]
\addplot [semithick, color0, const plot mark left, dashed,  mark size=3, mark options={solid}]
table {%
0 -6074.86239080703
1 -14224.2481962993
2 19532.7815340719
3 53138.8429297158
4 39017.9195281649
5 41720.8853107229
6 9076.4573308866
7 1671.95165809931
8 -4.76905166021245e-20
9 -4.76905166021245e-20
};
\addlegendentry{Company $a$}

\addplot [semithick, color1, const plot mark left, dashed,  mark size=3, mark options={solid}]
table {%
0 -2895.0999881975
1 -13732.9006650332
2 16947.7921897509
3 73363.9256390007
4 75830.480654844
5 49961.6101473778
6 6698.83593910272
7 1798.7128400984
8 -1.14683332104907e-19
9 -1.14683332104907e-19
};
\addlegendentry{Company $b$}

\addplot [semithick, black, const plot mark left, dashed,  mark size=3, mark options={solid}]
table {%
0 43.5406066378297
1 94.1821638520491
2 8393.27876229563
3 11009.1229777986
4 8655.11069917759
5 5463.44485883635
6 3290.04126805828
7 713.595950246939
8 5000
9 5000
};
\addlegendentry{Lost}

\end{axis}

\end{tikzpicture}
    \end{adjustbox}
    \caption{\textbf{Closed loop:} Attained profits for $T=6$.}
    \label{fig:profit6}
\end{figure}
\begin{figure}[!t]
    \centering
    \begin{adjustbox}{ max width=0.48\textwidth}
\begin{tikzpicture}

\definecolor{color0}{rgb}{1,0.498039215686275,0.0549019607843137}
\definecolor{color1}{rgb}{0.580392156862745,0.403921568627451,0.741176470588235}

\begin{axis}[
legend cell align={right},
legend style={fill opacity=0.8, draw opacity=1, text opacity=1, draw=white!90!black,at={(1,1)}, font=\small},
tick align=outside,
tick pos=left,
x grid style={white!90!black},
xlabel={Time interval $k$},
xmajorgrids,
xmin=-0.2, xmax=9.2,
xtick style={color=black},
y grid style={white!90!black},
ylabel=\textcolor{black}{$\mathbf{J}_{i,k}^{\text{market}}(\mathbf{Z})-\mathbf{J}_{i,k}^{\text{charge}}(\mathbf{Z})$},
ymin=-18571.2316110751, ymax=110000.0800394689,
ytick style={color=black},
height=4cm,
width=8cm,
]
\addplot [semithick, color0, const plot mark left, dashed,  mark size=3, mark options={solid}]
table {%
0 -4016.93712565774
1 -11520.5221355486
2 28883.8304003865
3 42537.5286278405
4 30625.396635154
5 -1593.66726278345
6 -520.248720576893
7 1775.66338690699
8 2144.21931554458
9 2144.21931554458
};
\addlegendentry{Company $a$}

\addplot [semithick, color1, const plot mark left, dashed,  mark size=3, mark options={solid}]
table {%
0 -2354.63016142854
1 -9927.96423697593
2 19719.108371694
3 36001.206683866
4 29206.5682018454
5 -938.345352423196
6 -449.273770687497
7 1877.13516780676
8 2141.62066826569
9 2141.62066826569
};
\addlegendentry{Company $b$}

\addplot [semithick, black, const plot mark left, dashed,  mark size=3, mark options={solid}]
table {%
0 43.0258569639394
1 83.8896589397414
2 7364.99213000077
3 58004.3308784978
4 63643.6659271414
5 99999.5142270233
6 19999.8379318014
7 510.488553320642
8 714.16001618973
9 714.16001618973
};
\addlegendentry{Lost}

\end{axis}

\end{tikzpicture}
    \end{adjustbox}
    \caption{\textbf{Closed loop:} Attained profits for $T=3$.}
    \label{fig:profit3}
\end{figure}}{}
\bibliographystyle{IEEEtran}
\bibliography{references.bib}
\end{document}